\documentclass[twocolumn]{aastex63}
\usepackage{amsmath}
\usepackage{empheq}
\usepackage{mathrsfs}
\usepackage{textcomp}
\usepackage{gensymb}
\usepackage{hyperref}
\usepackage{graphicx}
\usepackage[caption=false]{subfig}
\usepackage{multirow}
\usepackage{longtable}

\turnoffedit

\hypersetup{colorlinks, linkcolor={blue}, citecolor={blue}, urlcolor={blue}}  

\newcommand{\slope}{\mbox{$\Delta(g-r)$/$\Delta t$}}
\newcommand{\referee}[1]{#1}%\textbf{#1}}

\begin{document}

\title{ZTF Early Observations of Type Ia  Supernovae III: \\ Early-Time Colors as a Test for Explosion Models and Multiple Populations}

\correspondingauthor{Mattia Bulla}
\email{mattia.bulla@fysik.su.se}

\author[0000-0002-8255-5127]{Mattia Bulla}
\affiliation{Nordita, KTH Royal Institute of Technology and Stockholm University, Roslagstullsbacken 23, SE-106 91 Stockholm, Sweden}
\affil{The Oskar Klein Centre, Department of Physics, Stockholm University, AlbaNova, SE-10691 Stockholm, Sweden}

\author[0000-0001-9515-478X]{Adam A. Miller}
\affiliation{Center for Interdisciplinary Exploration and Research in Astrophysics (CIERA) and Department of Physics and Astronomy, Northwestern University, 2145 Sheridan Road, Evanston, IL 60208, USA}
\affiliation{The Adler Planetarium, Chicago, IL 60605, USA}

\author[0000-0001-6747-8509]{Yuhan Yao}
\affiliation{Division of Physics, Mathematics, and Astronomy, California Institute of Technology, Pasadena, CA 91125, USA}

\author{Luc Dessart}
\affiliation{Unidad Mixta Internacional Franco-Chilena de Astronom\'ia, CNRS/INSU UMI 3386
and Instituto de Astrof\'isica, Pontificia Universidad Cat\'olica de Chile, Santiago, Chile}

\author{Suhail Dhawan}
\affiliation{The Oskar Klein Centre, Department of Physics, Stockholm University, AlbaNova, SE-10691 Stockholm, Sweden}

\author[0000-0003-0783-3323]{Semeli Papadogiannakis}
\affiliation{The Oskar Klein Centre, Department of Physics, Stockholm University, AlbaNova, SE-10691 Stockholm, Sweden}

\author[0000-0002-5741-7195]{Rahul Biswas}
\affiliation{The Oskar Klein Centre, Department of Physics, Stockholm University, AlbaNova, SE-10691 Stockholm, Sweden}

\author[0000-0002-4163-4996]{Ariel Goobar}
\affiliation{The Oskar Klein Centre, Department of Physics, Stockholm University, AlbaNova, SE-10691 Stockholm, Sweden}

\author[0000-0001-5390-8563]{S. R. Kulkarni}
\affiliation{Division of Physics, Mathematics, and Astronomy, California Institute of Technology, Pasadena, CA 91125, USA}

\author{Jakob Nordin}
\affiliation{Institute of Physics, Humboldt-Universit\"at zu Berlin, Newtonstr. 15, 12489 Berlin, Germany}

\author[0000-0002-3389-0586]{Peter Nugent}
\affiliation{Departments of Physics and Astronomy, University of California, Berkeley, Berkeley, CA 94720 USA}
\affiliation{Lawrence Berkeley National Laboratory, Berkeley, CA 94720, USA}

\author{Abigail Polin}
\affiliation{Departments of Physics and Astronomy, University of California, Berkeley, Berkeley, CA 94720 USA}
\affiliation{Lawrence Berkeley National Laboratory, Berkeley, CA 94720, USA}

\author[0000-0003-1546-6615]{Jesper Sollerman}
\affiliation{The Oskar Klein Centre, Department of Astronomy, Stockholm University, AlbaNova, SE-10691 Stockholm, Sweden}

\author[0000-0001-8018-5348]{Eric C. Bellm}
\affiliation{DIRAC Institute, Department of Astronomy, University of Washington, 3910 15th Avenue NE, Seattle, WA 98195, USA} 

\author[0000-0002-8262-2924]{Michael W. Coughlin}
\affiliation{Division of Physics, Mathematics, and Astronomy, California Institute of Technology, Pasadena, CA 91125, USA}

\author{Richard Dekany}
\affiliation{Caltech Optical Observatories, California Institute of Technology, Pasadena, CA}

\author[0000-0001-8205-2506]{V. Zach Golkhou}
\altaffiliation{Moore-Sloan, WRF Innovation in Data Science, and DIRAC Fellow}
\affiliation{DIRAC Institute, Department of Astronomy, University of Washington, 3910 15th Avenue NE, Seattle, WA 98195, USA} 
\affiliation{The eScience Institute, University of Washington, Seattle, WA 98195, USA}

\author[0000-0002-3168-0139]{Matthew J. Graham}
\affiliation{Division of Physics, Mathematics, and Astronomy, California Institute of Technology, Pasadena, CA 91125, USA}

\author[0000-0002-5619-4938]{Mansi M. Kasliwal}
\affiliation{Division of Physics, Mathematics, and Astronomy, California Institute of Technology, Pasadena, CA 91125, USA}

\author[0000-0002-6540-1484]{Thomas Kupfer}
\affiliation{Kavli Institute for Theoretical Physics, University of California, Santa Barbara, CA 93106, USA}

\author[0000-0003-2451-5482]{Russ R. Laher}
\affiliation{IPAC, California Institute of Technology, 1200 E. California
             Blvd, Pasadena, CA 91125, USA}

\author[0000-0002-8532-9395]{Frank J. Masci}
\affiliation{IPAC, California Institute of Technology, 1200 E. California
             Blvd, Pasadena, CA 91125, USA}

\author{Michael Porter}
\affiliation{Caltech Optical Observatories, California Institute of Technology, Pasadena, CA}

\author[0000-0001-7648-4142]{Ben Rusholme}
\affiliation{IPAC, California Institute of Technology, 1200 E. California
             Blvd, Pasadena, CA 91125, USA}

\author[0000-0003-4401-0430]{David L. Shupe}
\affiliation{IPAC, California Institute of Technology, 1200 E. California
             Blvd, Pasadena, CA 91125, USA}

\begin{abstract}
Colors of Type Ia supernovae in the first few days after explosion provide a potential discriminant between different models. In this paper, we present $g-r$ colors of \edit1{65} Type Ia supernovae discovered within \edit1{5} days from first light by the Zwicky Transient Facility in 2018, a sample \referee{that} is \edit1{about three times larger than} that in the literature. We find that $g-r$ colors are intrinsically rather homogeneous at early phases, with about half of the dispersion attributable to photometric uncertainties \mbox{($\sigma_\mathrm{noise}\sim\sigma_\mathrm{int}\sim$~\edit1{0.18}~mag)}. Colors are nearly constant starting from \edit1{6}~days after first light \mbox{($g-r$~$\sim-0.15$~mag)}, while 
the time evolution at earlier epochs is characterized by a continuous range of slopes, from events rapidly transitioning from redder to bluer colors (slope of \mbox{$\sim-0.25$ mag day$^{-1}$}) to events with a flatter evolution. The continuum in the slope distribution is in good agreement both with models requiring some amount of $^{56}$Ni mixed in the outermost regions of the ejecta and with ``double-detonation'' models having thin helium layers ($M_\mathrm{He}=0.01\,M_\odot$) and varying carbon-oxygen core masses. At the same time, \edit1{six} events show evidence for a distinctive ``red bump'' signature predicted by ``double-detonation'' models with larger helium masses. We finally identify a significant correlation between the early-time $g-r$ slopes and supernova brightness, with brighter events associated to flatter color evolution (\edit1{p-value=0.006}). The distribution of slopes, however, is consistent with being drawn from a single population, with no evidence for two components as claimed in the literature \referee{based on $B-V$ colors}.
%Our color slopes are in conflict with the early-time evolution expected from the interaction of supernova ejecta with a non-degenerate companion star, posing serious challenges to this scenario for explaining the bulk of Type Ia supernovae.  
\end{abstract}

\section{Introduction}
\label{sec:intro}

Decades of observational and theoretical efforts have led to a general consensus that Type Ia supernovae (SNe~Ia) arise from thermonuclear explosions of carbon-oxygen white dwarfs in binary systems. Nevertheless, the conditions leading to the thermonuclear runaway are still debated, with the proposed scenarios typically grouped depending on whether the companion star is a non-degenerate star \citep[``single-degenerate channel'',][]{Whelan1973} or another white dwarf \citep[``double-degenerate channel'',][]{Iben1984,Webbink1984}, and whether the explosion mechanism is triggered close to the Chandrasekhar-mass ($M_\mathrm{ch}$) limit or in a sub-$M_\mathrm{ch}$ white dwarf (see \referee{e.g.,} \citealt{Livio2018} for a recent review).

Colors of SNe~Ia are controlled by the interplay between cooling from the ejecta expansion and heating due to thermalization of gamma-rays from the decay of radioactive elements (but are also affected by composition and line blanketing effects). Especially at early times, the color evolution can be used to probe the location within the ejecta of $^{56}$Ni and other radioactive isotopes \citep{Dessart2014} and help discriminate between different models. For instance, models producing $^{56}$Ni in the high-density innermost regions of the ejecta are expected to have red colors early on -- when the relatively cold outer ejecta are probed -- while showing a transition to bluer colors with the photosphere receding into increasingly hotter layers. In contrast, models with radioactive material mixed in the outer ejecta will be relatively bluer at early phases due to the additional source of heating from radioactive decay. 

An interesting example in this respect is the so-called sub-$M_\mathrm{ch}$ ``double-detonation'' scenario, where a first detonation in a thin helium layer accreted on the surface triggers a second detonation in the carbon-oxygen core \citep[\referee{e.g.,}][]{Nomoto1980,Taam1980,Livne1990,Fink2010,Moll2013}. Radioactive elements are produced both in a thin outermost layer and in the inner regions. These two distinct radioactive sources lead to blue colors at different times (soon after explosion and a few days later, respectively), with the transition in between producing a distinctive signature at early times, dubbed ``red bump'' \citep{Noebauer2017,Maeda2018,Polin2019a}. The ``double-detonation'' mechanism has been invoked to explain three recent SN Ia events \citep{Jiang2017,De2019,JacobsonGalan2019}. Other interesting scenarios involving the interaction of SN ejecta with either a non-degenerate companion star \citep{Kasen2010} or unbound material ejected prior to detonation \citep[pulsational-delayed-detonation models,][]{Dessart2014} predict rather blue colors soon after explosion ($g-r\lesssim0$~mag).

Early-time observations of SNe~Ia are challenging and thus have historically been limited to very nearby events. \citet{Stritzinger2018} presented a sample of 13 SNe~Ia \edit1{with colors at epochs earlier than 5 days} from inferred first light. Based on the $B-V$ color evolution in the first $\sim$~5 days, they claim evidence for two distinct populations, with a so-called ``red'' class showing a steep transition from red to bluer colors and a ``blue'' class characterized by bluer colors and flatter evolution. They suggested that events in the ``blue'' class are preferentially over-luminous and of the Branch Shallow Silicon (SS) spectral type, while those in the ``red'' class are more typically associated to the Branch Core-Normal (CN) or CooL (CL) type \citep{Branch2006}. Similar conclusions were drawn by \citet{Jiang2018} when inspecting light curves of 23 relatively young SNe~Ia. Recently, \citet{Han2020} added six events to the sample of \citet{Stritzinger2018} and claimed to confirm the distinction between ``red'' and ``blue'' events (but see discussion in Section~\ref{sec:pop}). 

Thanks to the advent of wide-field optical surveys, discovering SNe~Ia in their infancy has now become easier \citep[\referee{e.g.,}][]{Hosseinzadeh2017,Miller2018,Dimitriadis2019,Li2019,Papadogiannakis2019,Shappee2019,Vallely2019}. As the final in a series of three papers, here we report colors of \edit1{65}~SNe~Ia discovered within \edit1{5}~days from inferred first light by the Zwicky Tranient Facility \citep[ZTF,][]{bellm2019,graham2019,Masci2019} in 2018, a sample that to date is \edit1{about three times larger than} that available in the literature\footnote{Here we count only events \edit1{with the first color measurement within 5}~days from first light, \referee{i.e.,} a total of 19 SNe combining the sample of \citet{Stritzinger2018} and \citet{Han2020}. The sample of \citet{Jiang2018} reports discovery phases relative to maximum light rather than first light (see their table 1).}. In particular, we study the $g-r$ color evolution of our sample to place constraints on explosion models and at the same time test claims of two distinct populations in the early-time colors. Details of the sample are discussed in \citet{Yao2019}, while the analysis of $g$ and $r$ light-curves is presented in \citet{Miller2020}.

\begin{deluxetable*}{llcccccccc}[h]
%\tabletypesize{\scriptsize}
\tablecaption{Properties for the 65~SNe~Ia in our sample.}
\tablehead{
\colhead{ZTF Name}
& \colhead{TNS Name}
& \colhead{Ia Subtype} 
& \colhead{Redshift}
& \colhead{$t^\mathrm{first}_\mathrm{g-r}-t_\mathrm{fl}$} 
& \colhead{\slope} 
& \colhead{SALT2 $x_1$}
& \colhead{$E(B-V)_\mathrm{host}$}
& \colhead{$<K_{gr}>$}\\
%\colhead{(ZTF18)}   
& \colhead{}   
& \colhead{}
& \colhead{(days)}   
& \colhead{(mag day$^{-1}$)}   
& \colhead{}   
& \colhead{(mag)}
& \colhead{(mag)} \\
\colhead{(1)}  
& \colhead{(2)}  
& \colhead{(3)}  
& \colhead{(4)}  
& \colhead{(5)}  
& \colhead{(6)}  
& \colhead{(7)} 
& \colhead{(8)} 
& \colhead{(9)} 
}
\startdata
ZTF18aapqwyv & SN\,2018bhc & normal* & 0.0560 & 2.11$_{-0.69}^{+0.53}$ & -0.16 $\pm$ 0.16 & -1.72 $\pm$ 0.18 & 0.259 & 0.042 \\
ZTF18aapsedq & SN\,2018bgs & normal* & 0.0720 & 3.72$_{-0.31}^{+0.31}$ & - & -0.09 $\pm$ 0.18 & 0.011 & 0.073 \\
ZTF18aaqcugm & SN\,2018bhi & normal & 0.0619 & 4.50$_{-0.24}^{+0.22}$ & - & -1.12 $\pm$ 0.12 & 0.005 & 0.050 \\
ZTF18aaqqoqs & SN\,2018cbh & 99aa-like & 0.082 & 3.30$_{-0.23}^{+0.23}$ & - & 1.22 $\pm$ 0.27 & 0.044 & 0.083 \\
ZTF18aarldnh & SN\,2018lpd & normal & 0.1077 & 3.84$_{-0.64}^{+0.57}$ & -0.28 $\pm$ 0.25 & -1.05 $\pm$ 0.38 & 0.141 & 0.065 \\
ZTF18aasdted & SN\,2018big & normal & 0.0181 & 1.25$_{-0.10}^{+0.09}$ & - & 0.85 $\pm$ 0.05 & 0.257 & -0.001 \\
ZTF18aaslhxt & SN\,2018btk & normal & 0.0551 & 2.15$_{-0.09}^{+0.09}$ & - & 0.29 $\pm$ 0.02 & 0.000 & 0.039 \\
ZTF18aaumlfl & SN\,2018btg & normal & 0.0874 & 4.13$_{-0.40}^{+0.37}$ & - & -1.13 $\pm$ 0.26 & 0.027 & 0.070 \\
ZTF18aauocnw & SN\,2018cae & normal & 0.102 & 3.22$_{-0.57}^{+0.50}$ & -0.05 $\pm$ 0.14 & 0.14 $\pm$ 0.27 & 0.131 & 0.088 \\
ZTF18aavrwhu & SN\,2018bxo & normal & 0.0620 & 4.62$_{-0.22}^{+0.21}$ & - & 1.20 $\pm$ 0.27 & 0.044 & 0.060 \\
ZTF18aaxcntm & SN\,2018ccl & normal & 0.0269 & 3.78$_{-0.15}^{+0.15}$ & - & -1.52 $\pm$ 0.06 & 0.213 & 0.012 \\
ZTF18aaxdrjn & SN\,2018cdt & normal & 0.0340 & 4.47$_{-0.13}^{+0.13}$ & - & -1.92 $\pm$ 0.09 & 0.000 & 0.021 \\
ZTF18aaxqyki & SN\,2018cnz & normal & 0.1003 & 3.70$_{-0.54}^{+0.49}$ & - & 0.94 $\pm$ 0.27 & 0.025 & 0.075 \\
ZTF18aaxsioa & SN\,2018cfa & normal* & 0.0315 & 3.39$_{-0.07}^{+0.07}$ & 0.00 $\pm$ 0.02 & -1.51 $\pm$ 0.06 & 0.150 & 0.013 \\
ZTF18aaxvpsw & SN\,2018cof & normal & 0.0916 & 4.10$_{-0.56}^{+0.48}$ & -0.07 $\pm$ 0.12 & 0.04 $\pm$ 0.35 & 0.083 & 0.069 \\
ZTF18aaxwjmp & SN\,2018coe & normal & 0.084 & 4.04$_{-0.21}^{+0.20}$ & -0.03 $\pm$ 0.13 & 0.42 $\pm$ 0.15 & 0.100 & 0.081 \\
ZTF18aayjvve & SN\,2018cny & normal & 0.0474 & 2.19$_{-0.43}^{+0.38}$ & -0.13 $\pm$ 0.02 & -0.09 $\pm$ 0.10 & 0.164 & 0.036 \\
ZTF18aaykjei & SN\,2018crl & Ia-CSM & 0.0970 & 4.33$_{-0.31}^{+0.30}$ & - & 4.14 $\pm$ 0.21 & 0.000 & 0.090 \\
ZTF18aazblzy & SN\,2018cri & normal & 0.0653 & 1.36$_{-0.09}^{+0.08}$ & -0.06 $\pm$ 0.06 & -1.68 $\pm$ 0.09 & 0.016 & 0.054 \\
ZTF18aazixbw & SN\,2018coi & normal & 0.0594 & 2.67$_{-0.14}^{+0.13}$ & -0.05 $\pm$ 0.10 & -1.58 $\pm$ 0.13 & 0.147 & 0.054 \\
ZTF18aazsabq & SN\,2018crn & normal & 0.060 & 2.71$_{-0.63}^{+0.53}$ & -0.15 $\pm$ 0.01 & -1.24 $\pm$ 0.12 & 0.123 & 0.044 \\
ZTF18abatffv & SN\,2018lpf & normal & 0.143 & 4.45$_{-0.64}^{+0.57}$ & - & 0.95 $\pm$ 0.56 & 0.117 & 0.099 \\
ZTF18abauprj & SN\,2018cnw & 99aa-like & 0.0242 & 1.38$_{-0.10}^{+0.10}$ & -0.05 $\pm$ 0.02 & 1.34 $\pm$ 0.04 & 0.029 & -0.003 \\
ZTF18abaxlpi & SN\,2018ctm & normal & 0.0642 & 1.64$_{-0.17}^{+0.17}$ & -0.04 $\pm$ 0.04 & 0.14 $\pm$ 0.20 & 0.160 & 0.051 \\
ZTF18abcflnz & SN\,2018cuw & normal & 0.0273 & 2.82$_{-0.22}^{+0.20}$ & -0.12 $\pm$ 0.04 & 0.11 $\pm$ 0.02 & 0.050 & 0.000 \\
ZTF18abckujg & SN\,2018cvt & normal & 0.075 & 2.68$_{-0.25}^{+0.26}$ & - & 0.50 $\pm$ 0.30 & 0.078 & 0.062 \\
ZTF18abckujq & SN\,2018cvf & normal & 0.0638 & 3.07$_{-0.40}^{+0.37}$ & - & 0.51 $\pm$ 0.39 & 0.008 & 0.058 \\
ZTF18abclfee & SN\,2018cxk & 02cx-like & 0.0290 & 0.46$_{-0.16}^{+0.12}$ & 0.02 $\pm$ 0.01 & -2.53 $\pm$ 0.09 & 0.087 & 0.024 \\
ZTF18abcrxoj & SN\,2018cvw & normal & 0.0309 & 0.98$_{-0.07}^{+0.07}$ & -0.02 $\pm$ 0.03 & -1.29 $\pm$ 0.06 & 0.161 & 0.013 \\
ZTF18abdbuty & SN\,2018dbd & normal & 0.059 & 2.65$_{-0.26}^{+0.25}$ & -0.06 $\pm$ 0.03 & -0.76 $\pm$ 0.31 & 0.138 & 0.047 \\
ZTF18abdefet & SN\,2018dds & normal & 0.074 & 3.66$_{-0.81}^{+0.70}$ & 0.11 $\pm$ 0.06 & -0.12 $\pm$ 0.31 & 0.265 & 0.064 \\
ZTF18abdfydj & SN\,2018dzr & normal & 0.076 & 3.99$_{-0.30}^{+0.30}$ & - & 0.24 $\pm$ 0.26 & 0.054 & 0.079 \\
ZTF18abdkimx & SN\,2018dyq & normal & 0.077 & 4.00$_{-0.46}^{+0.43}$ & - & -0.05 $\pm$ 0.05 & 0.079 & 0.078 \\
ZTF18abdpvnd & SN\,2018dvf & SC & 0.050 & 3.58$_{-0.21}^{+0.21}$ & - & 3.06 $\pm$ 0.10 & 0.074 & 0.030 \\
ZTF18abeecwe & SN\,2018dje & normal & 0.0393 & 2.13$_{-0.09}^{+0.09}$ & -0.06 $\pm$ 0.04 & -0.56 $\pm$ 0.11 & 0.135 & 0.015 \\
ZTF18abeegsl & SN\,2018eag & normal & 0.072 & 4.12$_{-0.40}^{+0.37}$ & - & -2.20 $\pm$ 0.19 & 0.109 & 0.056 \\
ZTF18abetehf & SN\,2018dvb & normal & 0.0649 & 2.89$_{-0.14}^{+0.14}$ & -0.01 $\pm$ 0.09 & -1.37 $\pm$ 0.23 & 0.000 & 0.057 \\
ZTF18abfgygp & SN\,2018ead & normal & 0.064 & 2.66$_{-0.53}^{+0.45}$ & -0.02 $\pm$ 0.05 & 0.08 $\pm$ 0.02 & 0.037 & 0.059 \\
ZTF18abfhaji & SN\,2018dsw & normal & 0.084 & 2.90$_{-0.20}^{+0.19}$ & 0.02 $\pm$ 0.12 & -0.19 $\pm$ 0.04 & 0.056 & 0.072 \\
ZTF18abfhryc & SN\,2018dhw & normal & 0.0323 & 4.21$_{-0.59}^{+0.53}$ & - & 0.47 $\pm$ 0.04 & 0.084 & 0.003 \\
\enddata
\tablecomments{Column (3): classification from \citet{Yao2019}, ending with an asterisk in cases where classification could not be reliably determined from spectroscopy alone. Column (4): redshift from \citet{Yao2019}, shown with three decimals when inferred from \textsc{snid} fit of SN spectra and with four decimals otherwise. Column (5): rest-frame time of first detection in both $g$ and $r$ relative to first light $t_\mathrm{fl}$. Column (6): $g-r$ linear slope in the first \textbf{6} days for the \textbf{35} SNe with at least three data points in this time window. Column (7): SALT2 $x_1$ parameter from \citet{Yao2019}. Column (8): host reddening inferred using \texttt{SNooPy} \citep{Burns2014}. Column (9): averaged $K$-correction in the first 5~days since $t_\mathrm{fl}$ inferred using \texttt{SNooPy} \citep{Burns2014}.
}
\end{deluxetable*}

\begin{deluxetable*}{llcccccccc}[!t]
\ContinuedFloat
\tablecaption{Continued. \label{tab:parameters}}
%\tabletypesize{\scriptsize}
\tablehead{
\colhead{ZTF Name}
& \colhead{TNS Name}
& \colhead{Ia Subtype} 
& \colhead{Redshift}
& \colhead{$t^\mathrm{first}_\mathrm{g-r}-t_\mathrm{fl}$} 
& \colhead{\slope} 
& \colhead{SALT2 $x_1$}
& \colhead{$E(B-V)_\mathrm{host}$}
& \colhead{$<K_{gr}>$}\\
%\colhead{(ZTF18)}   
& \colhead{}   
& \colhead{}
& \colhead{(days)}   
& \colhead{(mag day$^{-1}$)}   
& \colhead{}   
& \colhead{(mag)}
& \colhead{(mag)} \\
\colhead{(1)}  
& \colhead{(2)}  
& \colhead{(3)}  
& \colhead{(4)}  
& \colhead{(5)}  
& \colhead{(6)}  
& \colhead{(7)} 
& \colhead{(8)} 
& \colhead{(9)} 
}
\startdata
ZTF18abfwuwn & SN\,2018ecq & 99aa-like* & 0.109 & 4.38$_{-0.22}^{+0.21}$ & - & 0.60 $\pm$ 0.30 & 0.059 & 0.080 \\
ZTF18abgmcmv & SN\,2018eay & 91T-like & 0.0185 & 1.20$_{-0.15}^{+0.14}$ & -0.06 $\pm$ 0.01 & 0.69 $\pm$ 0.05 & 0.770 & -0.009 \\
ZTF18abgxvra & SN\,2018efb & normal & 0.104 & 3.14$_{-0.20}^{+0.19}$ & 0.12 $\pm$ 0.07 & 0.80 $\pm$ 0.24 & 0.003 & 0.060 \\
ZTF18abimsyv & SN\,2018eni & normal* & 0.088 & 2.71$_{-0.15}^{+0.14}$ & 0.02 $\pm$ 0.06 & 1.05 $\pm$ 0.16 & 0.033 & 0.083 \\
ZTF18abjtger & SN\,2018err & normal & 0.107 & 4.79$_{-1.00}^{+0.91}$ & - & 0.77 $\pm$ 0.66 & 0.069 & 0.071 \\
ZTF18abjvhec & SN\,2018emv & normal & 0.0570 & 3.39$_{-0.31}^{+0.31}$ & - & 0.37 $\pm$ 0.41 & 0.052 & 0.052 \\
ZTF18abkhcrj & SN\,2018emi & normal & 0.0383 & 3.68$_{-0.22}^{+0.21}$ & -0.05 $\pm$ 0.05 & 0.85 $\pm$ 0.16 & 0.310 & 0.022 \\
ZTF18abkhcwl & SN\,2018eml & normal & 0.0317 & 3.46$_{-0.43}^{+0.39}$ & - & 0.08 $\pm$ 0.09 & 0.027 & 0.014 \\
ZTF18abkhdxe & SN\,2018ffg & normal & 0.104 & 4.55$_{-0.67}^{+0.62}$ & - & 0.73 $\pm$ 0.43 & 0.167 & 0.056 \\
ZTF18abmmkaz & SN\,2018fdz & 99aa-like* & 0.063 & 4.78$_{-2.87}^{+3.12}$ & - & 0.74 $\pm$ 0.15 & 0.028 & 0.047 \\
ZTF18abmxdhb & SN\,2018fjv & normal & 0.070 & 4.96$_{-0.43}^{+0.42}$ & - & 1.27 $\pm$ 0.22 & 0.026 & 0.056 \\
ZTF18abokpvh & SN\,2018fnc & normal* & 0.081 & 3.41$_{-0.21}^{+0.20}$ & 0.04 $\pm$ 0.06 & 0.77 $\pm$ 0.22 & 0.000 & 0.057 \\
ZTF18abpamut & SN\,2018fqe & normal* & 0.064 & 1.00$_{-0.51}^{+0.37}$ & -0.03 $\pm$ 0.00 & 0.83 $\pm$ 0.29 & 0.185 & 0.061 \\
ZTF18abpaywm & SN\,2018fne & normal & 0.040 & 1.70$_{-0.14}^{+0.14}$ & 0.11 $\pm$ 0.10 & 0.61 $\pm$ 0.14 & 0.309 & 0.019 \\
ZTF18abpmmpo & SN\,2018fnd & 99aa-like & 0.076 & 3.98$_{-0.25}^{+0.23}$ & - & 1.50 $\pm$ 0.27 & 0.034 & 0.064 \\
ZTF18abpttky & SN\,2018fse & normal & 0.084 & 3.92$_{-0.51}^{+0.45}$ & - & -1.31 $\pm$ 0.40 & 0.073 & 0.066 \\
ZTF18absdgon & SN\,2018frx & normal* & 0.0620 & 3.63$_{-0.21}^{+0.20}$ & -0.08 $\pm$ 0.06 & -0.26 $\pm$ 0.18 & 0.295 & 0.048 \\
ZTF18abssuxz & SN\,2018gfe & normal & 0.0649 & 2.45$_{-0.31}^{+0.32}$ & - & -1.14 $\pm$ 0.17 & 0.150 & 0.055 \\
ZTF18abukmty & SN\,2018lpz & normal* & 0.104 & 3.76$_{-0.42}^{+0.39}$ & 0.23 $\pm$ 0.25 & 0.51 $\pm$ 0.32 & 0.085 & 0.066 \\
ZTF18abvbayb & SN\,2018lpq & normal & 0.132 & 3.36$_{-0.36}^{+0.34}$ & 0.00 $\pm$ 0.14 & -0.20 $\pm$ 0.31 & 0.048 & 0.056 \\
ZTF18abwdcdv & SN\,2018gre & normal & 0.0538 & 2.50$_{-0.17}^{+0.17}$ & -0.07 $\pm$ 0.11 & -0.46 $\pm$ 0.12 & 0.457 & 0.042 \\
ZTF18abwnsoc & SN\,2018lpr & normal & 0.099 & 3.71$_{-0.28}^{+0.28}$ & 0.09 $\pm$ 0.07 & 0.34 $\pm$ 0.34 & 0.098 & 0.072 \\
ZTF18abwtops & SN\,2018lqa & normal & 0.101 & 3.78$_{-0.36}^{+0.34}$ & -0.21 $\pm$ 0.02 & -1.38 $\pm$ 0.28 & 0.015 & 0.055 \\
ZTF18abxxssh & SN\,2018gvj & normal & 0.0782 & 3.52$_{-0.21}^{+0.20}$ & - & 1.53 $\pm$ 0.24 & 0.000 & 0.061 \\
ZTF18abxygvv & SN\,2018gwb & normal* & 0.079 & 1.63$_{-0.16}^{+0.15}$ & -0.07 $\pm$ 0.11 & -0.10 $\pm$ 0.22 & 0.020 & 0.057 \\
\enddata
\end{deluxetable*}

The paper is organized as follows. We provide details of the sample selection and of the analysis in Section~\ref{sec:sample}, while presenting the inferred \mbox{$g-r$} colors in Section~\ref{sec:results}. We then compare our data to models in Section~\ref{sec:models} and test the presence of multiple populations in Section~\ref{sec:pop}. We finally discuss our results and draw conclusions in Section~\ref{sec:concl}.

\section{Data sample} \label{sec:sample}

For our study, we use high-quality $g_\mathrm{ZTF}$ and $r_\mathrm{ZTF}$ (hereafter $g$ and $r$) light curves of SNe~Ia discovered by ZTF in 2018. Details of the sample selection are discussed in \citet[][see their table~2]{Yao2019}. Briefly, 247 spectroscopically classified SNe~Ia were found by the high-cadence (6 epochs per night, 3$g$+3$r$) ZTF partnership survey in 2018. Among these, 127 SNe were discovered earlier than $-10$~days (in rest frame) relative to $B$-band peak brightness.
%, have sufficiently high number of observations in both $g$ and $r$ bands and reference images constructed at least 25~days before peak brightness and with no SN light included. 
\mbox{Forced-PSF} photometry performed by \citet{Yao2019} is used in this work for all the SNe in the sample. Following suggestions from \citet[][see their section~3.5]{Yao2019}, we remove observations with either high reduced chi-square statistics ($\chi^2_\nu>4$) or large baseline offset $C$ ($|C|>15$). This cut reduces the sample to 94 events.

In this paper, we are interested in studying colors of SNe Ia during the early phases following the explosion. 
%Because photons produced from the radioactive decay of $^{56}$Ni have to diffuse throughout the ejecta, a ``dark phase'' is known to exist in SNe Ia between explosion and observed first light \citep{Piro2013,Piro2014,Mazzali2014}. 
As in \cite{Stritzinger2018}, we choose to describe the color evolution of SNe in our sample with respect to the first-light epoch $t_\mathrm{fl}$, inferred by simultaneously fitting the early-time flux in both $g$ ($f_g$) and $r$ ($f_r$) band
\begin{equation}
f_i(t) = C + H[t_\mathrm{fl}]\,A_i\,(t - t_\mathrm{fl})^{\alpha_i}~~~~~i=g,r~~~,
\label{eq:flux}
\end{equation}
where $A_i$ is a scale factor, $t$ is the time, $\alpha_i$ is a power-law index and $H[t_\mathrm{fl}]$ is the heaviside step function ($H=0$ for $t<t_\mathrm{fl}$ and $H=1$ otherwise). In this work, we adopt first-light epochs $t_\mathrm{fl}$ from \citet{Miller2020}, which report values for two different set of models: one where an uninformative prior is assumed for $\alpha_i$ and one where $\alpha_g=\alpha_r=2$ (\referee{i.e.,} the t$^2$ model widely used in the literature, also known as ``fireball'' model, \citealt{Riess1999}). For each SN, we use the Deviance Information Criterion (DIC, \citealt{Spiegelhalter2002}) to choose what model better describes the early light curve and thus to select the corresponding $t_\mathrm{fl}$ value (see \citealt{Miller2020} for more details).   %Specifically, we restrict to events that have sufficiently good light-curve fits and use the Deviance Information Criterion (DIC) for choosing which $t_\mathrm{fl}$ value between their uninformative and t$^2$ models better describe the early rise (see Miller et al. in prep. for more details).

Here, we adopt the same cut made by \citet{Stritzinger2018} and restrict to SNe that \edit1{have the first color measurement within 5}~days from $t_\mathrm{fl}$. This leads to a sample of \edit1{65}~SNe~Ia, which comprises \edit1{56} normal SNe Ia, \edit1{six} over-luminous (91T-/99aa-like) SNe Ia, one ``super-Chandrasekhar'' SN, one ``Ia-CSM'' SN and one ``02cx-like'' SN according to the spectroscopic classification in \citet{Yao2019}. Table~\ref{tab:parameters} provides information about the \edit1{65}~SNe~Ia. As expected, SNe at higher compared to lower redshifts are discovered relatively later in their evolution. Specifically, the \edit1{21} events at $z\gtrsim0.08$ are all discovered in both $g$ and $r$ filters later than 2.5~days after $t_\mathrm{fl}$.

%When testing the presence of multiple populations in Section~\ref{sec:pop}, however, we will base our discussion on $N_\mathrm{boot}$ ``Bootstrap samples'' to account for the relatively large uncertainties on both $t_\mathrm{fl}$ ($\sim$~1 day) and $g-r$ colors (up to $\sim$~0.4~mag for some events at discovery). Specifically, each Bootstrap sample is constructed by re-sampling $t_\mathrm{fl}$ from the posteriors presented in Miller et al. (in prep) and $g-r$ colors from resampling $g$ and $r$ fluxes from a Normal distribution set by the corresponding flux uncertainties. For our analysis, we set $N_\mathrm{boot}=1000$.

In order to decrease the uncertainties on each data point, we average observations within the same night and then select 3$\sigma$ detections for our analysis. We then calculate $g-r$ colors for nights with detections in both $g$ and $r$. The following corrections are applied to $g$ and $r$ photometry before calculating the $g-r$ colors: (i) time-dilation correction; (ii) Milky-Way reddening correction; (iii) host-galaxy reddening correction; (iv) $K$-correction. Redshift and $E(B-V)_\mathrm{MW}$ values from table 3 of \citet{Yao2019} are used for step (i) and (ii), while the full light curves\footnote{\referee{SNe in our sample are observed for a median of $\sim$~80\% of the nights in the first 30 days since discovery (see \citealt{Yao2019} for more details on the light-curve sampling).} } are fit using the program \texttt{SNooPy} \citep{Burns2014} to infer $E(B-V)_\mathrm{host}$ and $K$-correction values for step (iii) and (iv). Host reddening and $K-$correction values are reported for each SN in Table~\ref{tab:parameters}.

The samples of \citet{Stritzinger2018} and \citet{Han2020} include only low-redshift SNe ($0.001\lesssim z \lesssim0.023$), while our sample extends to higher redshifts ($0.018\lesssim z \lesssim$~\edit1{0.143}) and it thus requires $K$-corrections. We note that $K$-corrections are not well-known in the first few days following the explosion. In particular, \texttt{SNooPy} estimates $K$-corrections by adopting the spectral template from \citet{Hsiao2007}, defined from 15~days before peak, and using an extrapolation at earlier epochs. Nevertheless, we find in Section~\ref{sec:colors} that our $g-r$ colors agree well with those from the low-redshift sample of \citet{Stritzinger2018}, thus giving us confidence about the $K$-corrections applied to our sample. In addition, we will base most of the discussion on the \textit{time evolution} (Section~\ref{sec:evolution}) rather than the absolute values (Section~\ref{sec:colors}) of colors as this choice is less sensitive to uncertainties on $K$-corrections.

\section{Results} \label{sec:results}

In the following, we present our results and discuss the inferred colors (Section~\ref{sec:colors}) and color evolution (Section~\ref{sec:evolution}) for the \edit1{65}~SNe~Ia in our sample.

\subsection{Colors} \label{sec:colors}

\begin{figure*}[t]
    \centering
    \includegraphics[width=1\textwidth]{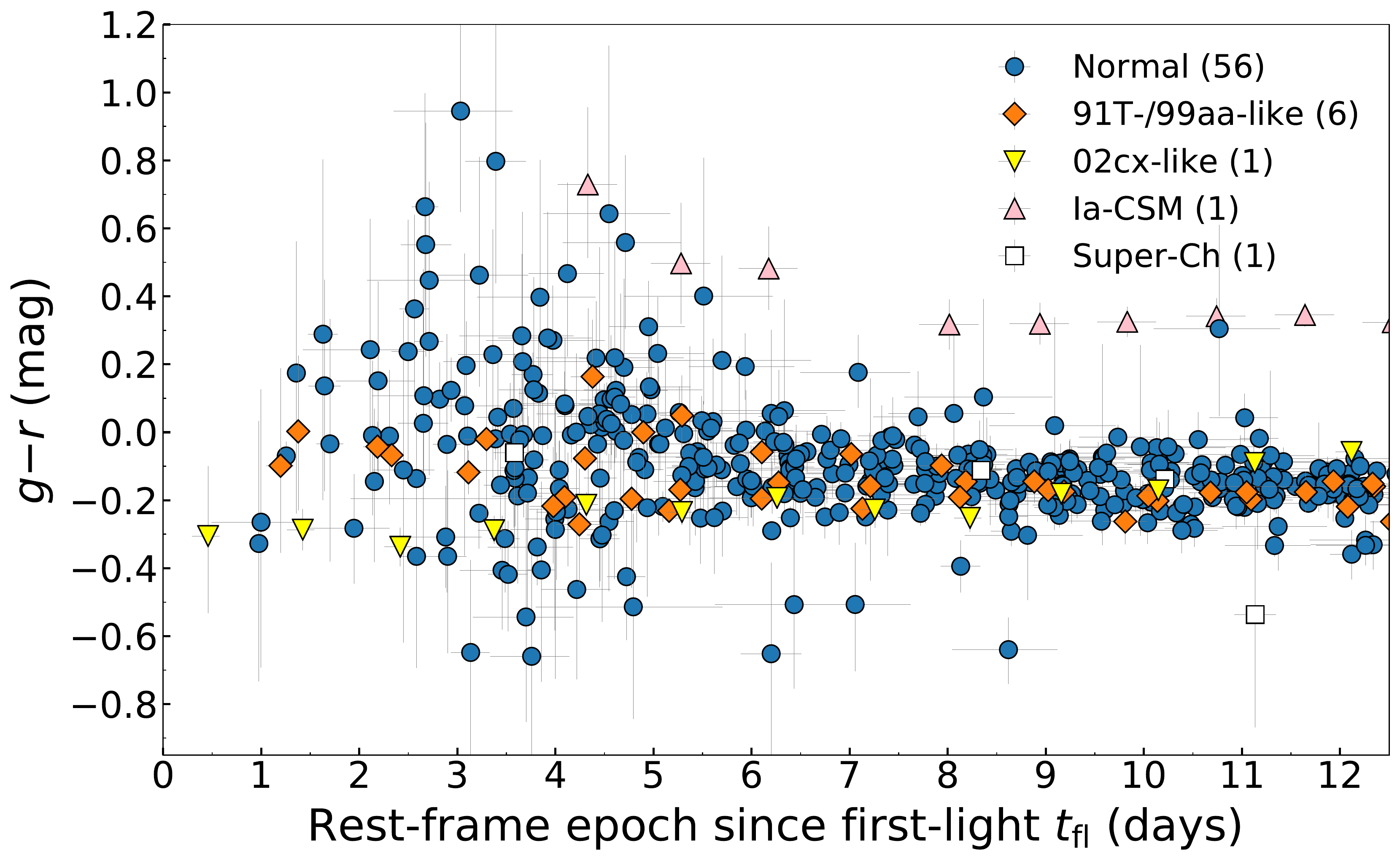}
   \caption{Evolution of $g-r$ colors for the \edit1{65}~SNe~Ia discovered by ZTF within \edit1{5}~days from first-light $t_\mathrm{fl}$. The sample includes \edit1{56} spectroscopically normal SNe~Ia (blue circles), \edit1{six} over-luminous 91T-/99aa-like SNe~Ia (orange diamonds), one ``02cx-like'' SN (yellow triangles down), one ``Ia-CSM'' SN (pink triangle up) and one ``super-Chandrasekhar'' SN (white squares). Colors are corrected for reddening (both Milky Way and host) and $K$-correction.  \label{fig:gr}}
\end{figure*}

\begin{figure}[t]
    \centering
    \includegraphics[width=1\columnwidth]{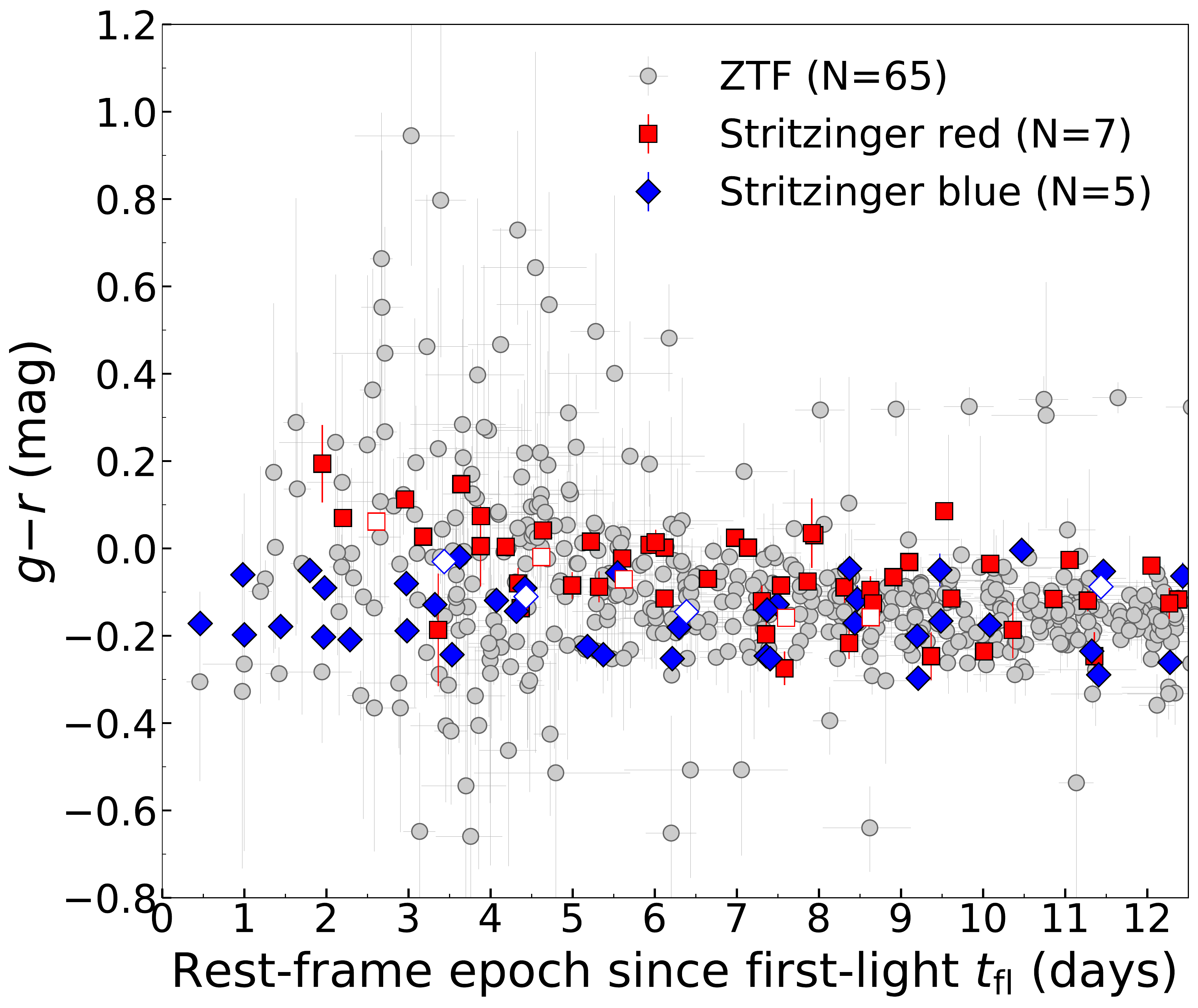}
   \caption{$g-r$ color evolution of our ZTF sample (grey circles), compared to that of 12 SNe~Ia from \citet{Stritzinger2018} that have available $g$ and $r$ photometry \citep[filled symbols,][]{Graham2015,Graham2017,Hsiao2015,Shappee2016,Hosseinzadeh2017,Burns2018,Miller2018,Vinko2018} or early-time spectra \citep[open symbols,][]{Foley2012,Silverman2012}. Following \citet{Stritzinger2018}, the 12 SNe are divided in ``red" (red squares) and ``blue" (blue diamonds) objects. Time of first-light and reddening values are taken from \citet{Stritzinger2018}. \label{fig:gr_stritz}}
\end{figure}

\begin{figure}[t]
    \centering
    \includegraphics[width=1\columnwidth]{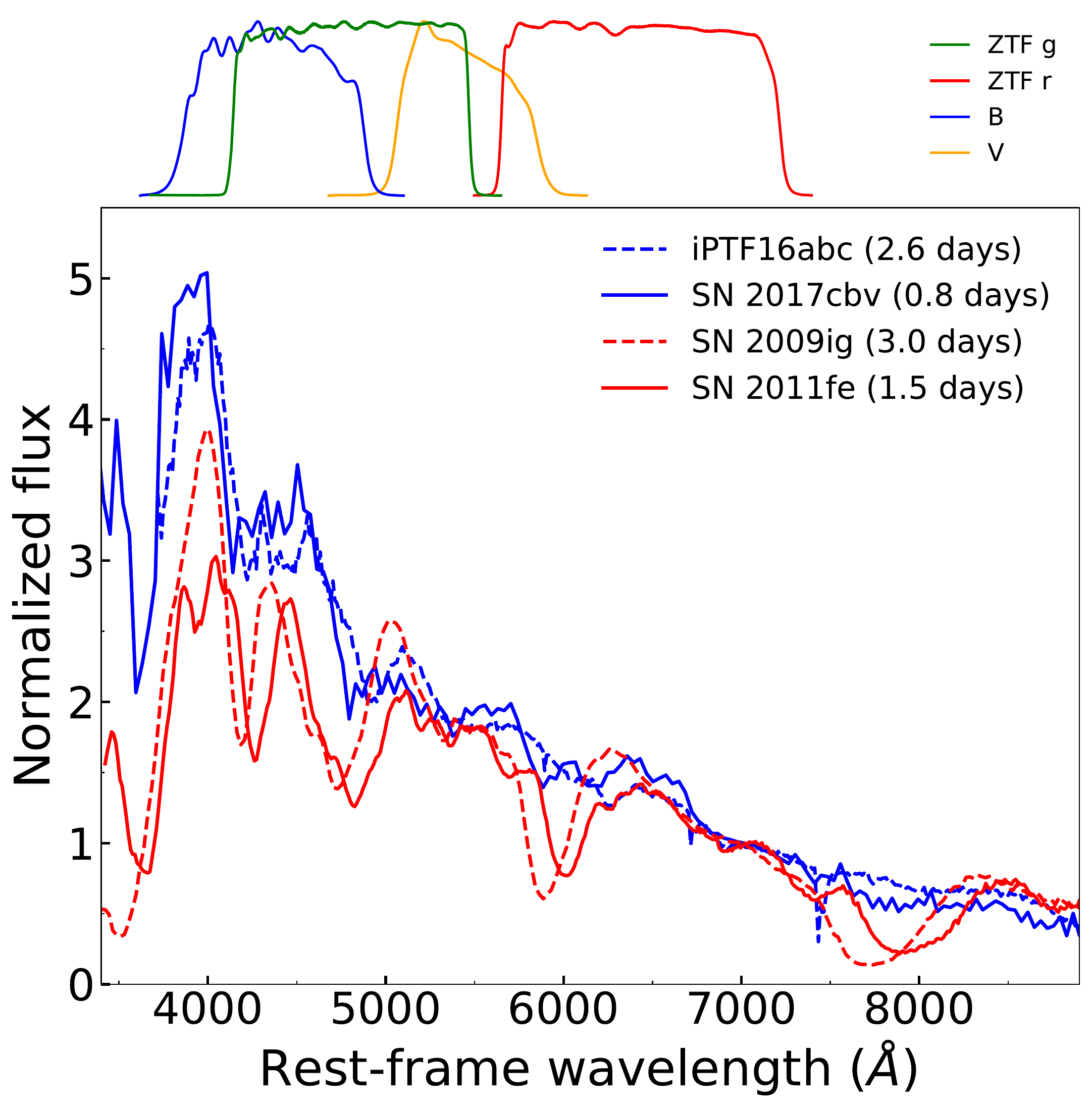}
   \caption{Early-time spectra of iPTF16abc \citep{Miller2018}, SN 2017cbv \citep{Hosseinzadeh2017}, SN 2009ig \citep{Foley2012} and SN 2011fe \citep{Nugent2011}. ZTF $g$ and $r$ filters are shown at the top together with $B$ and $V$ filters. Spectra have been normalized at 7000~\AA{} and rebinned for presentation purpose. \label{fig:spec_filt}}
\end{figure}

Figure~\ref{fig:gr} shows the $g-r$ colors for the sample of \edit1{65}~SNe~Ia discovered by ZTF within \edit1{5}~days from first light $t_\mathrm{fl}$. The distribution of $g-r$ values is rather homogeneous starting from about \edit1{6}~days after $t_\mathrm{fl}$, with colors clustering around $g-r \sim -0.15$~mag\footnote{The peculiar ``Ia-CSM'' SN ZTF18aaykjei (SN\,2018crl) is characterized by redder colors ($g-r\sim0.3$~mag) due to $H_\alpha$ emission at wavelengths covered by the ZTF $r$ filter.}. In contrast, the scatter is found to be larger at earlier epochs. However, some fraction of the scatter observed at very early times is caused by the relatively high photometric uncertainties that characterize most of our SNe when first detected. In particular, the typical uncertainties at these early epochs have a median value of $\sigma_\mathrm{noise}\sim$~\edit1{0.18}~mag, while the $g-r$ distribution in the first \edit1{6}~days after $t_\mathrm{fl}$ has a width of $\sigma$~=~\edit1{0.23}~mag. Following the light-curve rise and corresponding increase in signal-to-noise, both the uncertainties and the scatter in colors decrease, with the latter always $\sim40\,\%$ larger than the former. Based on these numbers, we conclude that roughly half of the scatter observed in our colors at early times ($\lesssim$~10~days) is intrinsic and half is due to photometric uncertainties, \referee{i.e.,} $\sigma_\mathrm{int}\sim\sigma_\mathrm{noise}\sim\sigma/\sqrt{2}$. The fact that $\sigma_\mathrm{int}\lesssim$~\edit1{0.18}~mag in the first \edit1{six} days after first light suggests that SNe~Ia are intrinsically more homogeneous in $g-r$ compared to what has been found in $B-V$ colors \citep{Stritzinger2018}. This is in qualitative agreement with the finding in \citet[][see their section 4.3]{Miller2020}.

The larger homogeneity of $g-r$ relative to $B-V$ colors is confirmed when comparing our sample to the 12 SNe~Ia from \citet{Stritzinger2018} that have available $g$ and $r$ photometry \citep{Graham2015,Graham2017,Hsiao2015,Shappee2016,Hosseinzadeh2017,Burns2018,Miller2018,Vinko2018} or early-time spectra to compute synthetic photometry \citep{Foley2012,Silverman2012}. As shown in Figure~\ref{fig:gr_stritz}, no clear gap is found in $g-r$ at early phases between the ``red'' and ``blue'' class introduced in $B-V$ colors by \citet{Stritzinger2018}, corroborating the idea that the early-time color evolution in SNe~Ia might be rather homogeneous in $g$ and $r$ filters. Figure~\ref{fig:gr_stritz} also highlights how the color evolution of the ZTF sample is consistent with that reported by \citet{Stritzinger2018}. The good agreement between the two samples gives us confidence about both the extinction- and $K$-corrections applied to our sample.

As shown in Figure~\ref{fig:spec_filt}, the larger homogeneity in $g-r$ compared to $B-V$ colors can be understood as a consequence of the different parts of the SED probed by different filter combinations. Early-time spectra of four SNe~Ia in the \citet{Stritzinger2018} sample are shown, where two events (iPTF16abc and SN 2017cbv) are from the so-called ``blue'' class and two (SN 2009ig and SN 2011fe) are from the ``red'' class. In the wavelength region probed by the four filters, the largest spectral diversities between the two classes are seen at wavelengths below $\sim$~4800~\AA{} and around the Si\,{\sc ii}~$\lambda6355$. These follow from ``blue'' objects being 91T-/99aa-like SNe, events that have been shown \citep[\referee{e.g.,}][]{Jeffery1992,Ruizlapuente1992,Mazzali1995} to be more highly ionized than normal SNe~Ia and thus lack singly-ionized absorption features such as Si\,{\sc ii}~$\lambda6355$ at these early phases. The ZTF $g$ and $r$ filters are broader than $B$ and $V$ filters and cover both regions with large spectral diversity. In contrast, while the $B$ filter probes the region below $\sim$~4800~\AA{}, the $V$ filter covers a region around 5000~\AA{} that is relatively homogeneous between the two classes. In addition, the $B$ filter extends to bluer wavelengths than the $g$ filter, in a spectral range ($\sim$~$3800-4200$~\AA) with pronounced differences between ``blue'' and ``red'' objects. Therefore, the largest contrast between the two classes is captured by $B-V$ colors, while $g-r$ colors tend to wash out the observed spectral differences (this is similar to what is found at later epochs by \citealt{Nordin2018}, see top panel of their figure 3). This comparison explains why $g-r$ colors are found to be more homogeneous than $B-V$ in the first few days after explosion. At the same time, it suggests that $B$ and $V$ filters might be the better choice to test different models affecting the early-time colors.

\subsection{Color evolution} \label{sec:evolution}

\begin{figure*}[t]
    \centering
    \includegraphics[width=\textwidth]{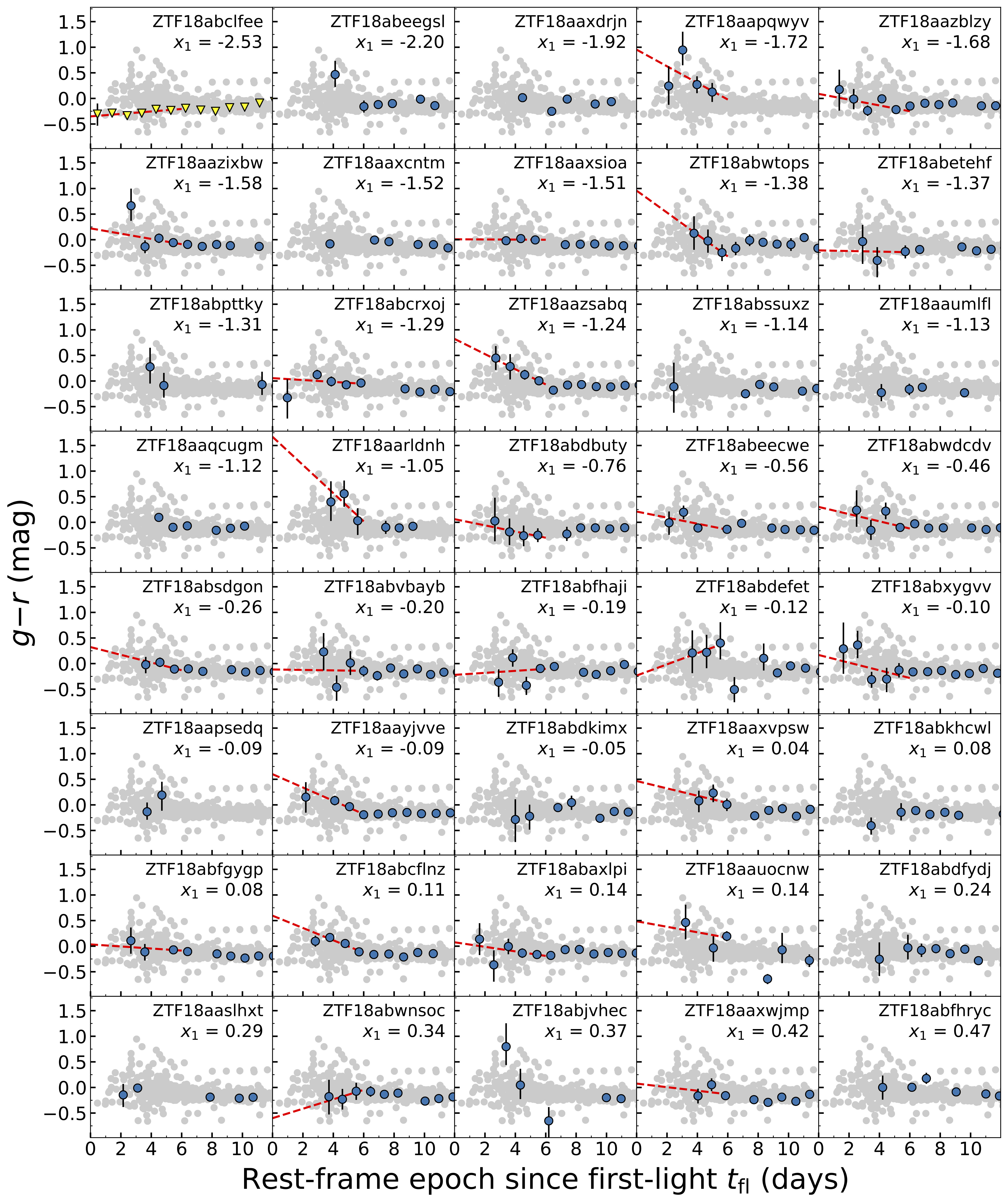}
   \caption{Same as Figure~\ref{fig:gr} but with the $g-r$ evolution of each of the \edit1{65} SNe highlighted. SNe are ordered from top-left to bottom-right according to their SALT2~$x_1$ values \citep{Yao2019}. Grey points mark the colors of the full sample for comparison. The red dashed line in each panel is a weighted least-square linear fit to colors in the first \edit1{6} days for events with at least three data points in this time window. Colors and symbols are the same as in Figure~\ref{fig:gr}. \label{fig:gr_ind}}
\end{figure*}

\begin{figure*}[t]
    \ContinuedFloat
    \centering
    \includegraphics[width=\textwidth]{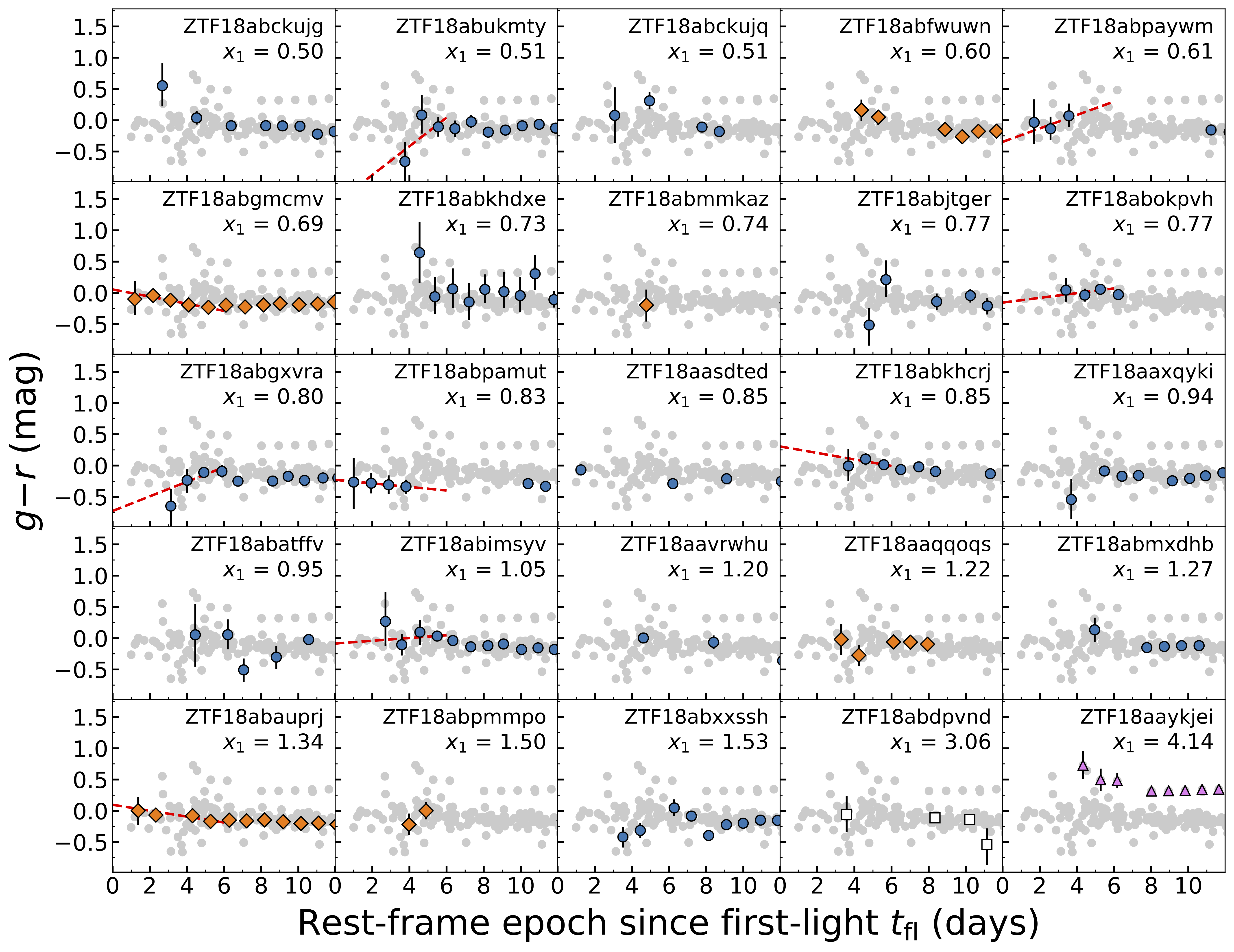}
   \caption{Continued.}
\end{figure*}

\begin{figure*}[t]
    \centering
    \includegraphics[width=\textwidth]{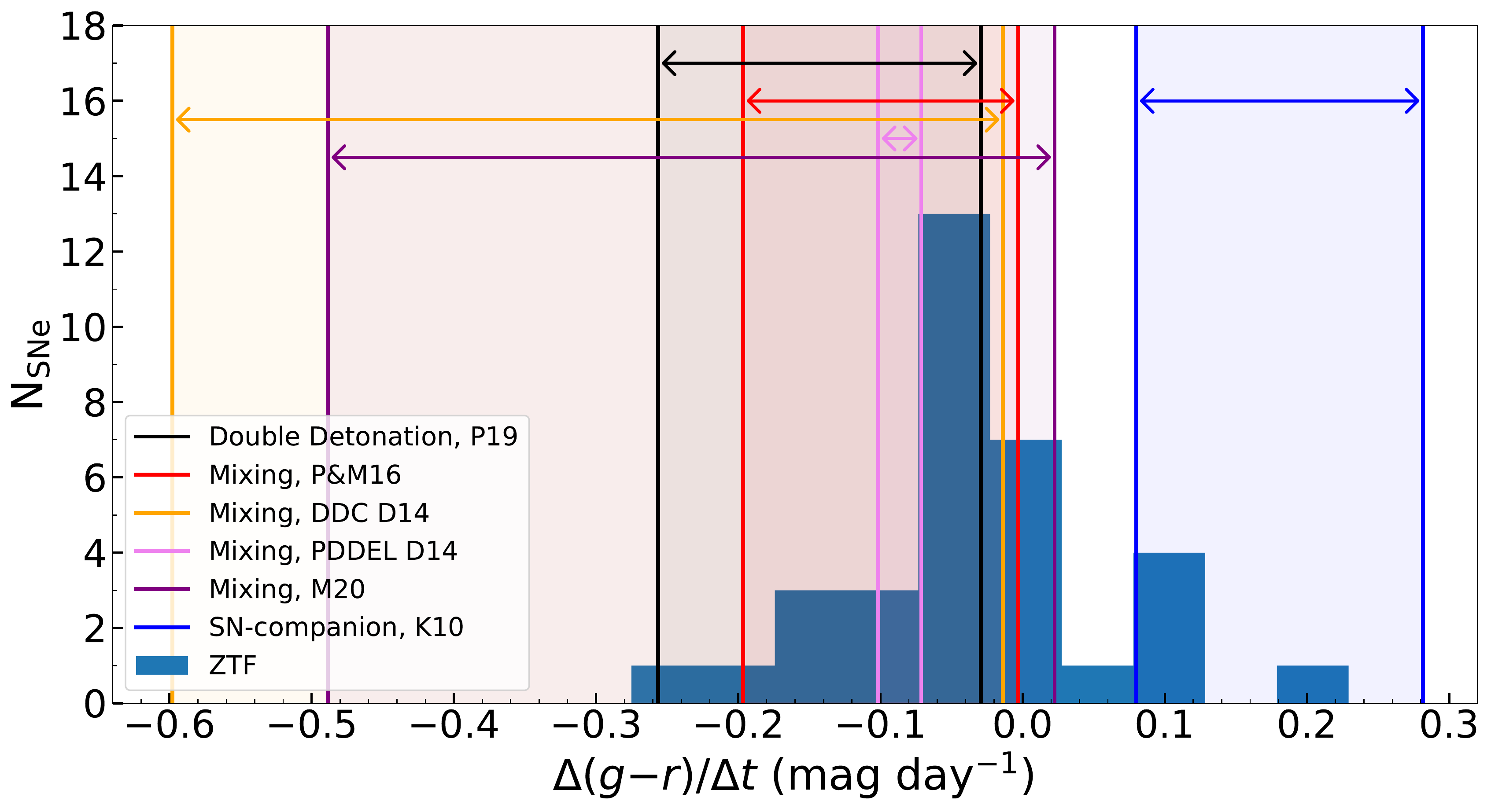}
    \caption{Comparison between observed and modelled slopes in the first \edit1{6}~days since first light. The distribution refers to the linear slopes \slope~measured for \edit1{34} SNe~Ia with at least three detections in the first \edit1{6}~days (see Section~\ref{sec:evolution}, the peculiar 02cx-like SN ZTF18abclfee/SN\,2018cxk is not considered here). The range spanned by each series of models is shown with a shaded area and with an horizontal arrow. Mixing models are from \citet[][red]{Piro2016}, \citet[][orange and violet]{Dessart2014} and \citet[][purple]{Magee2020}, with an increasing amount of mixing from left to right (vertical lines). The range spanned by the four SN-companion interaction models from \citet{Kasen2010} is shown in blue, while that from ``double-detonation'' models of \citet{Polin2019a} with $M_\mathrm{He}=0.01\,M_\odot$ in black. See text for more details.
   \label{fig:slope_models} }
\end{figure*}

Despite the homogeneity of $g-r$ values discussed above, we do see a distinct color evolution. Figure~\ref{fig:gr_ind} shows the $g-r$ color evolution of each individual SN in our sample. To characterize the change in colors we restrict ourselves to events %that are discovered within 2~days from $t_\mathrm{fl}$ and 
that have at least three data-points in the first \edit1{6}~days, resulting in a sample of \edit1{35}~SNe~Ia. We then characterize the color evolution by performing a weighted least-square linear fit to $g-r$ in the first \edit1{6}~days and infer a slope \slope~for each SN, with positive (negative) values associated to colors becoming redder (bluer). Results of these fits are shown in Figure~\ref{fig:gr_ind} and reported in Table~\ref{tab:parameters} for the \edit1{35}~SNe~Ia that meet the criteria defined above.

%adopting two different approaches:
%\begin{itemize}
%    \item \textit{Linear fit}: we perform a least-square linear fit to $g-r$ in the first 4~days and infer a slope \slope~for each SN, with positive (negative) values associated to colors becoming redder (bluer). Results of these fits are shown in Figure~\ref{fig:gr_ind} and reported in Table~\ref{tab:parameters} for the 18~SNe~Ia that meet the criteria defined above.
%    \item \textit{Power-law indexes}: we adopt the best-fit $\alpha_g$ and $\alpha_r$ from Miller et al. (in prep) and from Equation~\ref{eq:flux} derive the instantaneous slope in color
%    \begin{equation}
%        (g-r)^\prime=\frac{2.5}{\log10}\frac{\alpha_r-\alpha_g}{t-t_\mathrm{fl}}~~.
%        \label{eq:slope}
%    \end{equation}
%    In this approach, the color evolution is controlled by the difference in power law indexes \mbox{$\Delta\alpha=\alpha_r-\alpha_g$}, with $\Delta\alpha\sim0$ characterizing flat evolution and positive (negative) $\Delta\alpha$ values associated to colors becoming redder (bluer).
%\end{itemize}

As shown in Figure~\ref{fig:gr_ind}, some events are characterized by a clear transition from redder to bluer colors and thus a negative slope, \slope~$<$~0, others by a flatter evolution, \slope~$\sim$~0. We note that all the three over-luminous 91T-/99aa-like SNe are characterized by relatively flat color evolutions, in agreement with findings from \citet{Stritzinger2018}. The full range of slopes, going from a minimum of \slope~$\sim$~$-$0.28 to a maximum of \slope~$\sim$~\edit1{0.23}~mag day$^{-1}$, is reported in Figure~\ref{fig:slope_models}. The range in color evolution is reminiscent of the two classes introduced by \citet{Stritzinger2018}, with negative slopes consistent with their ``red'' class while flatter slopes with their ``blue'' class. When comparing data to models (Section~\ref{sec:models}) and when investigating the possible presence of multiple populations (Section~\ref{sec:pop}), we will focus on the time evolution \slope~
%~and $\Delta\alpha$ 
rather than the absolute values of colors. We consider this choice more robust as it is less sensitive to uncertainties introduced by both reddening corrections and $K$-corrections. 

\section{Comparison to models} \label{sec:models}

In this section, we compare the $g-r$ evolution of our sample with model predictions. In particular, we focus on three different scenarios that have been shown to produce characteristic signatures in the colors at early times (see discussion in Section~\ref{sec:intro}). Specifically, we explore the SN ejecta-companion interaction model (Section~\ref{sec:companion}), the ``double-detonation'' scenario (Section~\ref{sec:doubledet}) and models with different amounts of $^{56}$Ni mixed throughout the ejecta (Section~\ref{sec:mixing}). The peculiar ``02cx-like'', ``Ia-CSM'' and ``Super-Chandrasekhar'' events are not considered in these comparisons.

We note that models presented here are plotted relative to explosion, while data are shown relative to first light $t_\mathrm{fl}$. Many of the SNe~Ia in our sample (especially those at low redshift) are detected 4 to 5~mag below peak \citep{Yao2019} and thus $t_\mathrm{fl}-t_\mathrm{exp}$ is expected to be small for these events according to predictions from explosion models \citep[$\lesssim 2$~days, see \referee{e.g.,} figure 4 in][]{Dessart2014}. 
%Nevertheless, there could be some ``dark phases'' (see Section~\ref{sec:intro}) between explosion epoch, $t_\mathrm{exp}$, and $t_\mathrm{fl}$ (especially in low-mixing models, see Section~\ref{sec:mixing}). 
Nevertheless, given the issues with inferring $t_\mathrm{exp}$ from observations and with having a common definition of $t_\mathrm{fl}$ across different models, we choose not to apply any shift to either models or data but caution against making a one-to-one comparison between them.

\subsection{SN ejecta-companion interaction}
\label{sec:companion}

\begin{figure}[t]
    \centering
    \includegraphics[width=\columnwidth,clip=True,trim=8 0 0 0 0]{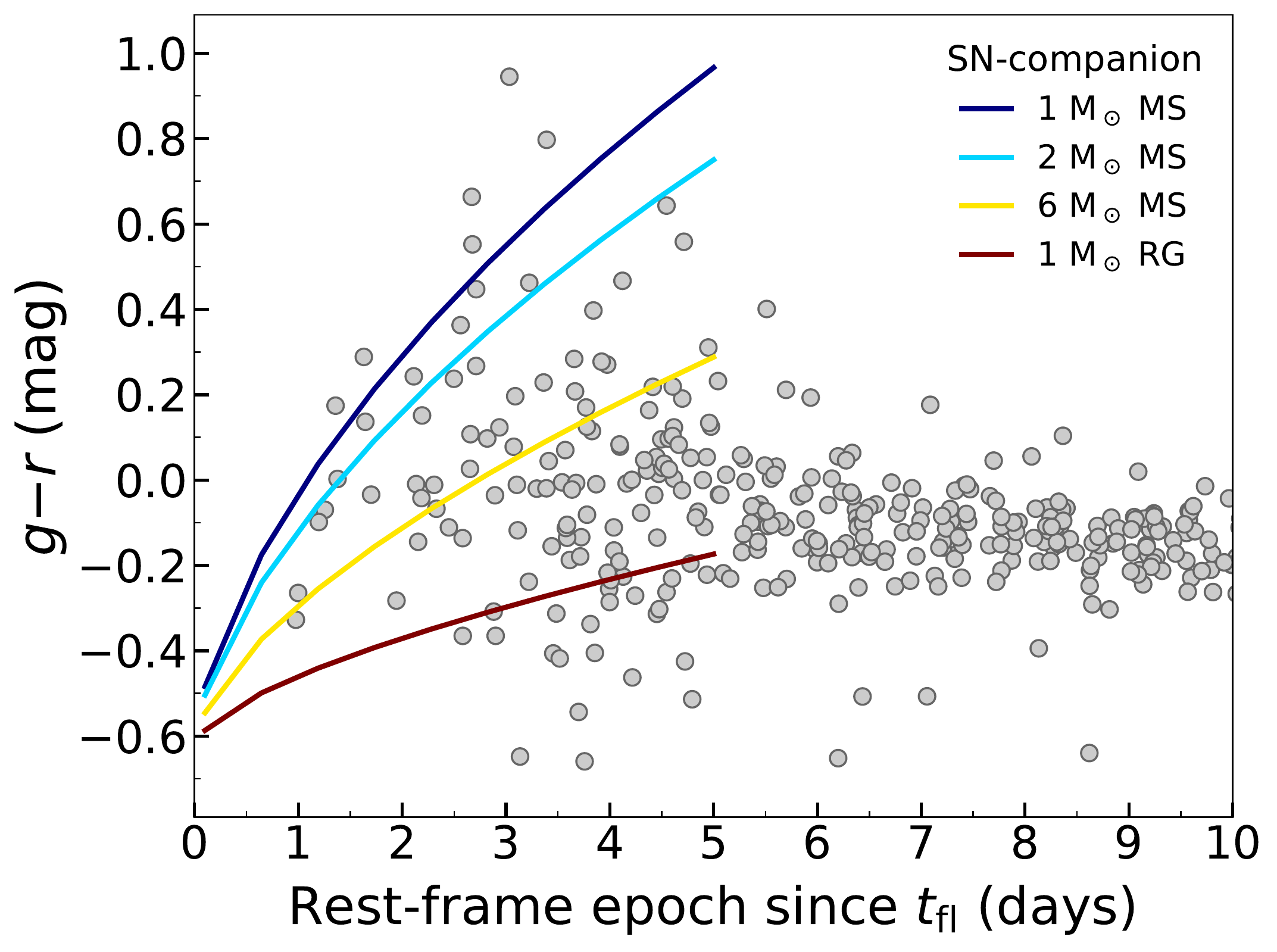}
   \caption{Comparison of our ZTF sample (grey points) to SN ejecta-companion models from \citet{Kasen2010}. Model predictions are shown for MS stars with three different masses (1, 2 and 6~$M_\odot$) and for a 1~$M_\odot$ RG star. Predicted colors are shown only in the first $\sim$~5 days since $t_\mathrm{fl}$ when the emission from the SN ejecta-companion interaction is expected to be dominant \citep[see \referee{e.g.,} equation 23 in][]{Kasen2010}. \label{fig:sncomp}}
\end{figure}

Figure~\ref{fig:sncomp} compares our sample to SN ejecta-companion models from \citet{Kasen2010}. Predicted colors are shown for the four different companion-star models discussed in \citet{Kasen2010}, \referee{i.e.,} three MS stars with different masses (1, 2 and 6~$M_\odot$) and a 1~$M_\odot$ RG star. Luminosity and temperature for each model is estimated using equation~22 and 25 in \citet{Kasen2010} and assuming an ejecta velocity $v=10^4$~km~s$^{-1}$ and an effective opacity $\kappa_\mathrm{e}=0.2$~cm$^2$~g$^{-1}$. Fluxes and corresponding $g-r$ colors are then estimated under a blackbody approximation. Curves are shown only in the first $\sim$~5 days since first light when the emission from the SN ejecta-companion interaction is expected to be dominant \citep[see \referee{e.g.,} equation 23 in][]{Kasen2010,Maeda2018}.

All the models investigated predict similar and relatively blue colors at first light, $g-r\sim-0.5$~mag, which then become redder with time following the decrease in temperature (see equation~25 in \citealt{Kasen2010}). The transition from bluer to redder colors is characterized by \slope~$\gtrsim0.1$~mag~day$^{-1}$, with a slower evolution in the case the companion is a RG or for increasing masses in the MS case. As shown in Figure~\ref{fig:slope_models} and summarized in Table~\ref{tab:parameters}, we see evidence for such a rapid transition in \edit1{five} events: ZTF18abgxvra (SN\,2018efb), \edit1{ZTF18abukmty (SN\,2018lpz)}, \edit1{ZTF18abwnsoc (SN\,2018lpr)}, \edit1{ZTF18abdefet (SN\,2018dds)} and ZTF18abpaywm (SN\,2018fne). However, the latter \edit1{four events are} characterized by relatively high photometric uncertainties (see Figure~\ref{fig:slope_models}) while \edit1{ZTF18abgxvra shows} a sign of ``red bump'' in \edit1{the} color evolution and might thus come from a ``double-detonation'' explosion (see Section~\ref{sec:doubledet} and right panel of Figure~\ref{fig:ddet}). Moreover, a good match in color slopes is found only with the 1~$M_\odot$ RG companion star model, which predicts a very strong bump in both UV and optical light curves \citep[see figure 3 of][]{Kasen2010} that is not found in any of these \edit1{five} events.
%, with a change in color of \slope~$\lesssim0.1$ mag day$^{-1}$ for all the 22~SNe for which we measure a slope (see Section~\ref{sec:evolution}, the peculiar 02cx-like SN ZTF18abclfee is not considered here).

%Flatter color evolutions as the one predicted by a 1~$M_\odot$ RG companion star might be compatible with the bluest objects in our sample. However, these models predict a very strong bump in both UV and optical light curves, which is not seen in any SN in our sample (Miller et al. in prep).

Our calculations assume a perfect alignment between the exploding white dwarf, the companion star and the observer. As shown by \citet{Kasen2010}, the signature of the collision should be prominent $\sim$~10$\%$ of the times for a favourable observer orientation near the perfect alignment. While we cannot exclude the presence of a companion star for each individual SN, the large size of our sample suggests we should see the effect of an interaction in $\sim$~\edit1{six} events. As a consequence, the fact that we do not see any clear evidence for a SN ejecta-companion interaction poses challenges to this scenario to explain the bulk of the SN Ia population. %At face value, the non-detection of a rapid blue-to-red transition in 32 events suggests that the SN ejecta-companion interaction might be rare and occur in $\lesssim3~\%$ of the cases.

\subsection{Helium-ignited Double Detonation models}
\label{sec:doubledet}

\begin{figure*}[t]
    \centering
    \includegraphics[width=\textwidth,clip=True,trim=8 0 0 0 0]{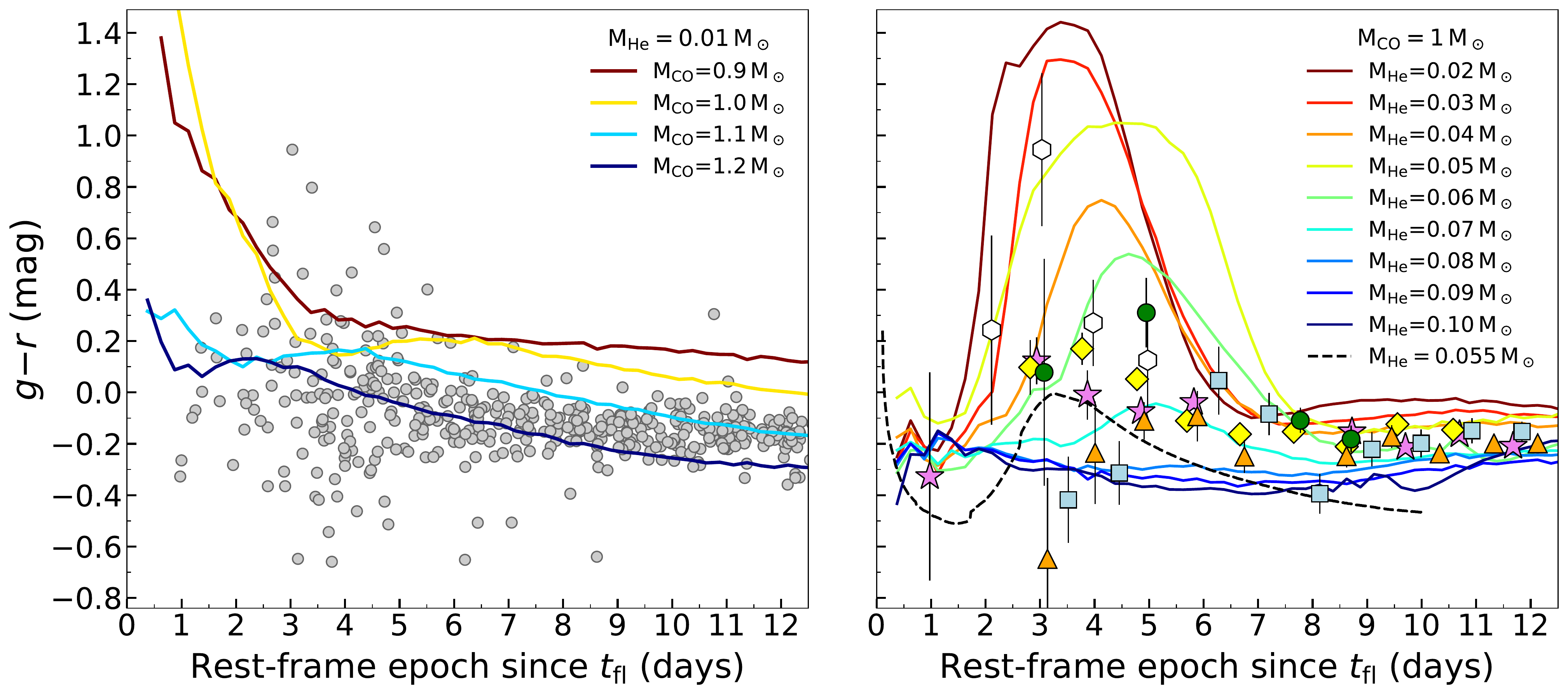}
   \caption{Comparison of our ZTF sample to ``double-detonation'' models. \textit{Left panel}: \referee{the full ZTF sample (grey points) compared to} models from \citet{Polin2019a} with fixed helium mass $M_\mathrm{He}=0.01\,M_\odot$ and carbon-oxygen mass varying in the range $M_{\rm CO}\in[0.9,1.2]\,M_\odot$. \textit{Right panel}: models with fixed carbon-oxygen mass $M_{\rm CO}=1.0\,M_\odot$ and varying helium masses. Models from \citet[][solid lines]{Polin2019a} have helium masses varying in the range $M_{\rm He}\in[0.02,0.10]\,M_\odot$ (from dark red to dark blue), while the model from \citet[][black dashed line]{Fink2010} as computed by \citet{Noebauer2017} has $M_{\rm He}=0.055\,M_\odot$. The \edit1{six} events \referee{in the ZTF sample} showing possible ``red bumps'', ZTF18abcflnz (SN\,2018cuw), ZTF18abxxssh (SN\,2018gvj), ZTF18abcrxoj (SN\,2018cvw), ZTF18abgxvra (SN\,2018efb), \edit1{ZTF18abckujq (SN\,2018cvf) and ZTF18aapqwyv (SN\,2018bhc)} are highlighted with yellow diamonds, light-blue squares, violet stars, orange triangles, \edit1{green circles and white hexagons,} respectively. \label{fig:ddet}}
\end{figure*}

Figure~\ref{fig:ddet} shows the $g-r$ evolution of our \edit1{65} SNe~Ia compared to that predicted by helium-ignited ``double-detonation'' models from the literature. The left panel includes models from \citet{Polin2019a} with fixed helium mass \mbox{$M_\mathrm{He}=0.01\,M_\odot$} and varying carbon-oxygen core masses, while the right panel models from \citet{Noebauer2017} and \citet{Polin2019a} with carbon-oxygen core masses of $M_{\rm CO}\sim1.0\,M_\odot$ and
%since they synthesise a $^{56}$Ni mass that can account for the brightness observed in normal SNe~Ia \citep{Stritzinger2006,Scalzo2014,Childress2015,Dhawan2016}. Models from \citet[][solid lines]{Polin2019a} have 
varying helium shell masses in the range $M_{\rm He}\in[0.02,0.10]\,M_\odot$.
%while the model of \citet{Fink2010} studied by \citet[][dashed line]{Noebauer2017} have $M_{\rm He}=0.055M_\odot$.

Models with very thin helium layers (left panel) show a range in the early-time color slopes. Models with $M_{\rm CO}=0.9$ and $1.0\,M_\odot$ are characterized by steep transitions from red to bluer colors, while those with $M_{\rm CO}=1.1$ and $1.2\,M_\odot$ by flatter evolutions. As shown in Figure~\ref{fig:slope_models}, this range in slopes is in reasonable agreement with that observed in our ZTF sample although it can not explain the events with \slope~$\gtrsim0$~mag~day$^{-1}$. We note that the four ``double-detonation'' models used here are those that have been claimed to explain maximum-light colors, velocity \citep{Polin2019a}, polarization \citep{Cikota2019} and nebular calcium emission \citep{Polin2019b} of a subset of SNe~Ia. Our findings bring additional support to these claims, suggesting that the ``double-detonation'' scenario might contribute to some fraction of the observed SN Ia population. Specifically, the comparison in Figure~\ref{fig:slope_models} suggests that the ``double detonation'' models can explain the range in slopes observed for $\sim$~60$\%$ (\edit1{21} out of \edit1{34}) of the events.

Models with relatively thicker helium layers ($0.02\lesssim M_{\rm He}\lesssim0.07\,M_\odot$, right panel) produce strong ``red bumps'' (see Section~\ref{sec:intro}). Visually inspecting the color evolution of each SN in Figure~\ref{fig:gr_ind}, we find evidence for a modest ``red bump'' in \edit1{six} events: ZTF18abcflnz (SN\,2018cuw), ZTF18abxxssh (SN\,2018gvj), ZTF18abcrxoj (SN\,2018cvw), ZTF18abgxvra (SN\,2018efb), \edit1{ZTF18abckujq (SN\,2018cvf) and ZTF18aapqwyv (SN\,2018bhc)}. All these \edit1{six} events display $g-r$ colors that are relatively blue at detection\footnote{We note that this statement relies somewhat on the rather large uncertainties in $g-r$ colors at detection.}, evolve to redder colors, reach $g-r \sim$~0 at $\sim3-$\edit1{6} days after $t_\mathrm{fl}$ and then turn over to bluer colors. This temporal evolution is in good agreement with predictions from \referee{e.g.,} \citet{Noebauer2017}, suggesting that these SNe might come from ``double-detonation'' explosions of sub-$M_\mathrm{ch}$ white dwarfs with relatively thick helium layers ($M_\mathrm{He}\sim0.05\,M_\odot$). In addition, ZTF18abxxssh (SN\,2018gvj) is characterized by a strong light-curve excess at early times \citep{Yao2019}, making the interpretation of this SN within the ``double-detonation'' framework even more viable. The detection in \edit1{6} out of \edit1{65} SNe suggests a ``red bump'' might occur in $\sim$~\edit1{9}$\%$ of the cases. 
%This number increases to $\sim$9$\%$ if including only spectroscopically normal SNe~Ia. 
We note that these estimates are not representative of the ``double-detonation'' contribution to the SN~Ia population, but rather of a subclass with relatively thick helium mass and thus detectable ``red bump''. As discussed above, ``double-detonation'' models with thin helium layers (\mbox{$M_\mathrm{He}=0.01\,M_\odot$}) might instead explain a good fraction ($\sim$~60$\%$) of the observed population.

\subsection{$^{56}$Ni mixing}
\label{sec:mixing}

\begin{figure}[htb]
    \centering
    \includegraphics[width=0.915\columnwidth,clip=True,trim=8 0 8 0]{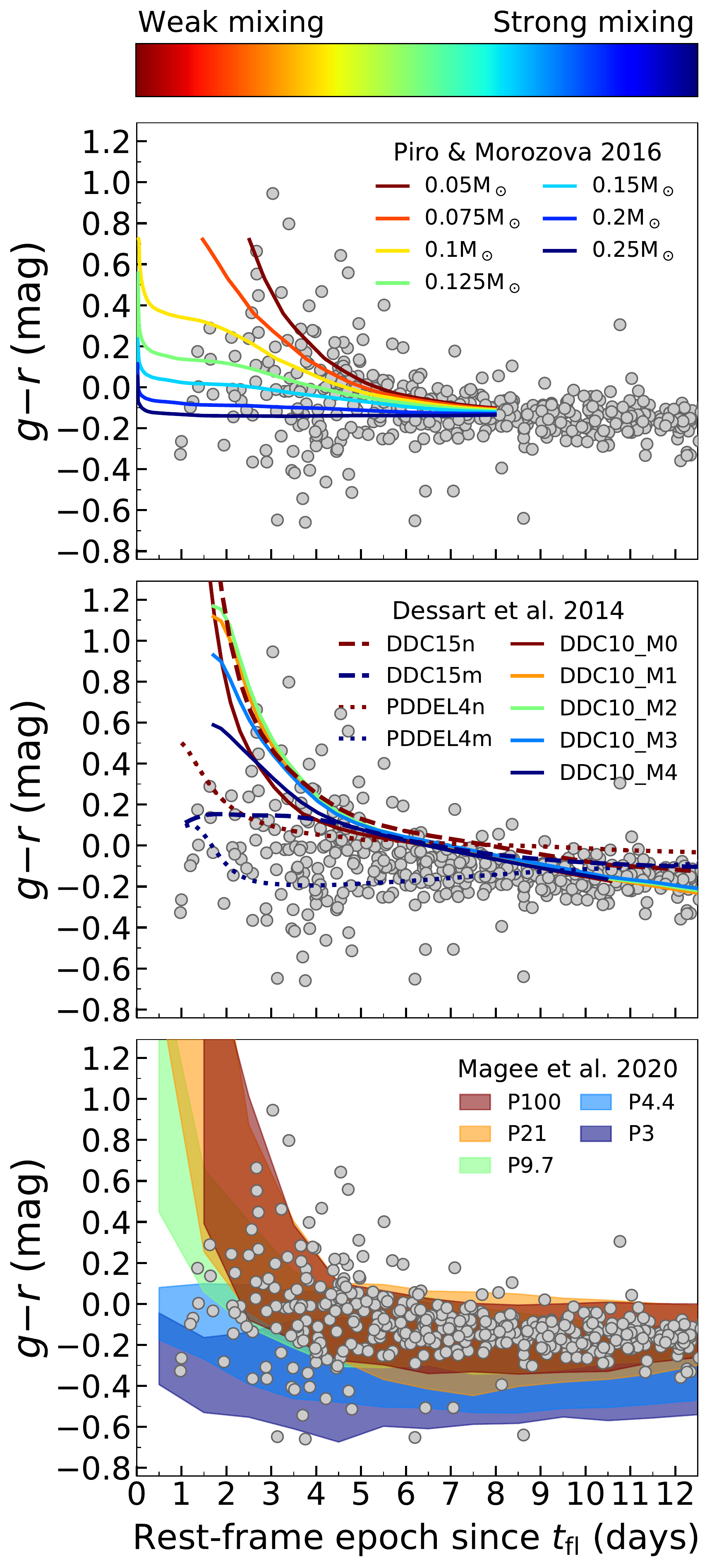}
   \caption{Comparison between our sample (grey points) and mixing models. The amount of mixing increases from models in red to models in blue. Top panel: models from \citet{Piro2016}. Middle panel: delayed-detonation (DDC, solid and dashed lines) and pulsational delayed detonation (PDDEL, dotted lines) models from \citet{Dessart2014}\referee{, together with the unpublished models DDC15m and PDDEL4m}. Bottom panel: models from \citet{Magee2020} using the radiative transfer code \textsc{turtls} \citep{Magee2018}. For each mixing model, the shaded area represents color variations for different density profile shapes and kinetic energies (see text for details). \label{fig:mixing}}
\end{figure}

%\begin{figure*}[t]
%    \centering
%    \includegraphics[width=0.95\textwidth]{plots/gr_lcpar.pdf}
%   \caption{Light-curve shape parameters $x_1$ against $g-r$ slopes, d$(g-r)$/d$t$, in the first 4 days after first light. The analysis is restricted to a sample of 32 events discovered earlier than 2~days from $t_\mathrm{fl}$ and with at least three measured colors earlier than 4~days (right panel). Colors and symbols are the same as in Figure~\ref{fig:gr}. The vertical dashed line marks the median value of d$(g-r)$/d$t$ and divides the sample between events with steeper (``Steep'') and flatter (``Flat'') color evolutions. The horizontal dashed line corresponds to $x_1=0$, dividing the sample in ``Bright'' and ``Faint'' events. Left panels show median $x_1$ values for the ``Steep'' and ``Flat'' class (top) and median d$(g-r)$/d$t$ slopes for the ``Bright'' and ``Faint'' class (bottom). The vertical lines refer to values obtained using the  ``fiducial sample'' shown in the right panel, while shaded distributions those for the 1000 ``Bootstrap samples''. The analysis suggests a possible association between blue and over-luminous (low $\Delta\mathrm{m}_{15}$(B)) events and between red and under-luminous events.
%   \label{fig:gr_lcpar}%}
%\end{figure*}

Figure~\ref{fig:mixing} shows comparison between our sample and models exploring different amounts of $^{56}$Ni mixing, where the color coding in all the different panels spans from red to blue for an increasing amount of mixing. 

The top panel refers to models of \citet{Piro2016}, where mixing is implemented using a ``boxcar'' average with widths between 0.05 and 0.25~$M_\odot$. 
%While configurations with strong mixing have negligible ``dark phases'' (see Section~\ref{sec:intro}), first light occurs $\sim$~2.5 and $\sim$~1.5~days after explosion in the 0.05 and 0.075~$M_\odot$ models, respectively. To properly compare simulations to data, we therefore plot models relative to first-light rather than to explosion epoch. 
As described in Section~\ref{sec:intro}, models with stronger mixing are characterized by bluer colors at early times and relatively flatter evolution. Models by \citet{Piro2016} are qualitatively in good agreement with our data, both in terms of colors and color evolution (see Figure~\ref{fig:slope_models}). This comparison tentatively suggests that some amount of mixing is required to reproduce the average colors in the first few days after first light. We note, however, that Local Thermodynamic Equilibrium (LTE) is assumed by \citet{Piro2016} and thus predicted colors should be treated with caution.
%. Predicted colors should thus be treated with caution as the opacities in the outer ejecta could be sufficiently high to reprocess light from shorter to longer optical wavelengths and thus make the overall colors redder \citep[see e.g.][]{Polin2019a}.

The middle panel of Figure~\ref{fig:mixing} shows comparison with models by \citet{Dessart2014} and \referee{a} more recent (and unpublished) \referee{incarnation} (\referee{DDC15m, this model was} computed using the same approach as in \citealt{Dessart2014} and \referee{differs} only in the strength of mixing, as explained below). Unlike in \citet{Piro2016}, \citet{Dessart2014} carry out radiative transfer calculations for hydrodynamical models of $M_\mathrm{ch}$ delayed-detonations (denoted as DDC10 and DDC15). All elements are mixed using a boxcar algorithm adopting a characteristic velocity $v_\mathrm{mix}=$~250 (DDC10$\_$M1), 500 (DDC10$\_$M2), 1000 (DDC10$\_$M3) and 1500\,km~s$^{-1}$ (DDC10$\_$M4). We also include the delayed-detonation model DDC15 \citep{Dessart2014}, characterized by a relatively weak mixing of elements (model DDC15n; $v_\mathrm{mix}=$~400\,km~s$^{-1}$). In contrast, \referee{the new unpublished} model DDC15m is strongly mixed and similar to the most mixed of the \citet{Piro2016} models (top panel of Figure~\ref{fig:mixing}). In model DDC15m, the mixing is done using $m_\mathrm{mix}=$~0.25\,$M_\odot$, together with a gaussian smoothing with a characteristic width of 300\,km~s$^{-1}$. These models also predict bluer and flatter colors for increasing amount of mixing, however, the colors in the first few days are relatively redder than those by \citet{Piro2016}\footnote{We note that the discrepancy could be reduced with a shift of $\sim1-2$~days to account for the difference between $t_\mathrm{fl}$ and $t_\mathrm{exp}$, see above.}. This is caused in part by the fact that the mixing in mass space pollutes the outer (high velocity) ejecta layers much more efficiently that mixing in velocity space. This arises because little mass is contained in the high velocity layers of the ejecta (in model DDC10, there is about 0.2\,$M_\odot$ beyond 15000\,km\,s$^{-1}$). There may also be an opacity effect. Line blanketing below 5000\,\AA\ remains strong out to large velocities well above the optical photosphere, so that the SN optical color is only set at large velocity. Guessing the SN color at the photosphere by inspecting the local LTE temperature is inaccurate and likely overestimates the true optical color. The strongly mixed model DDC15m is about 0.15\,mag redder than the most mixed model from Piro \& Morozova, and appears somewhat too red relative to the observed mean $g-r$ color distribution (see also \citealt{Dessart2014} and \citealt{Miller2018}). Although the colors are relatively redder than those observed, we note that the spread in slope predicted by the DDC10 and DDC15 suggests that some amount of mixing is required to explain the observed distribution shown in Figure~\ref{fig:slope_models}.

Also included in the middle panel of Figure~\ref{fig:mixing} are the pulsational delayed detonation models of \citet{Dessart2014}. The explosion mechanism in this scenario is similar to the delayed-detonation mechanism but here a delay is introduced between the initial deflagration and the subsequent detonation \citep{Hoeflich1996}. This first pulse partially unbinds the outer layers of the $M_\mathrm{ch}$ white dwarf, so that the delayed detonation leads to a strong interaction between the detonated inner ejecta and the marginally unbound outer ejecta. The interaction leads to a strong dissipation of kinetic energy into heat, the formation of a dense shell at around 10000 to 15000\,km\,s$^{-1}$, with little mass beyond\footnote{The pulsational detonation scenario may correspond to an explosion configuration similar to the merging of two white dwarfs followed by a detonation. The marginally bound material from the pulsation in the PDDEL model corresponds now to the material that was flung during the merger and created a cocoon around the detonating residual.}. \citet{Dessart2014} demonstrated that the early boost of the outer ejecta temperature had observable consequences for days on the luminosity and color, yielding brighter and bluer SNe. The models in \citet{Dessart2014} were however characterized by a weak mixing. Here, we recomputed the model PDDEL4 of \citet{Dessart2014} by using the same mixing recipe as for model DDC15m above. We refer to this model as PDDEL4m. For comparison, we include the weakly mixed model PDDEL4 (here called PDDEL4n) of \citet{Dessart2014}. As can be seen from Figure~\ref{fig:mixing}, model PDDEL4m yields much bluer colors with a flatter evolution than the delayed detonation model DDC15m (i.e., with no pulsation). Because of the red-to-blue transition predicted in the first $\sim$~3 days, this model struggles to reproduce the flatter-end of the observed \slope~distribution (see Figure~\ref{fig:slope_models}).

The bottom panel of Figure~\ref{fig:mixing} includes mixing models from \citet{Magee2020}, computed using the radiative transfer code \textsc{turtls} \citep{Magee2018}. The grid of light-curve models is constructed with varying four main parameters: the $^{56}$Ni mass (0.4, 0.6, 0.8~$M_\odot$), the density profile shape (double power law or exponential), the kinetic energy and the amount of $^{56}$Ni mixing (see \citealt{Magee2020} for more details). Here we compare our data to models producing 0.6~$M_\odot$ of $^{56}$Ni and for each mixing value plot the range covered by different density profiles and kinetic energy. The comparison highlights how the observed $g-r$ evolution is well reproduced by models requiring some degree of $^{56}$Ni mixing (see Figure~\ref{fig:slope_models}). In particular, the strongest agreement with data in the first \edit1{6}~days is found for the ``P100'', ``P21'', ``P9.7'' and ``P4.4'' mixing models, with $\sim$~\edit1{67}\% of the data-points falling in the color range predicted by these models. We note that the more stratified models ``P100'' and ``P21'' were disfavoured by \citet{Magee2020} based on comparisons to early light curves of normal SNe Ia.

Mixing is parametrized in all the models presented above and thus discrepancies with data do not necessarily rule out mixing scenarios but perhaps suggest that the mixing is different than adopted. Nevertheless, the range in slopes measured for our sample is in good agreement with the color evolution predicted by mixing models and better explained by incarnations requiring relatively strong $^{56}$Ni mixing throughout the ejecta.

\section{Testing for multiple populations} \label{sec:pop}

\begin{figure*}[t]
    \centering
    \includegraphics[width=0.98\textwidth]{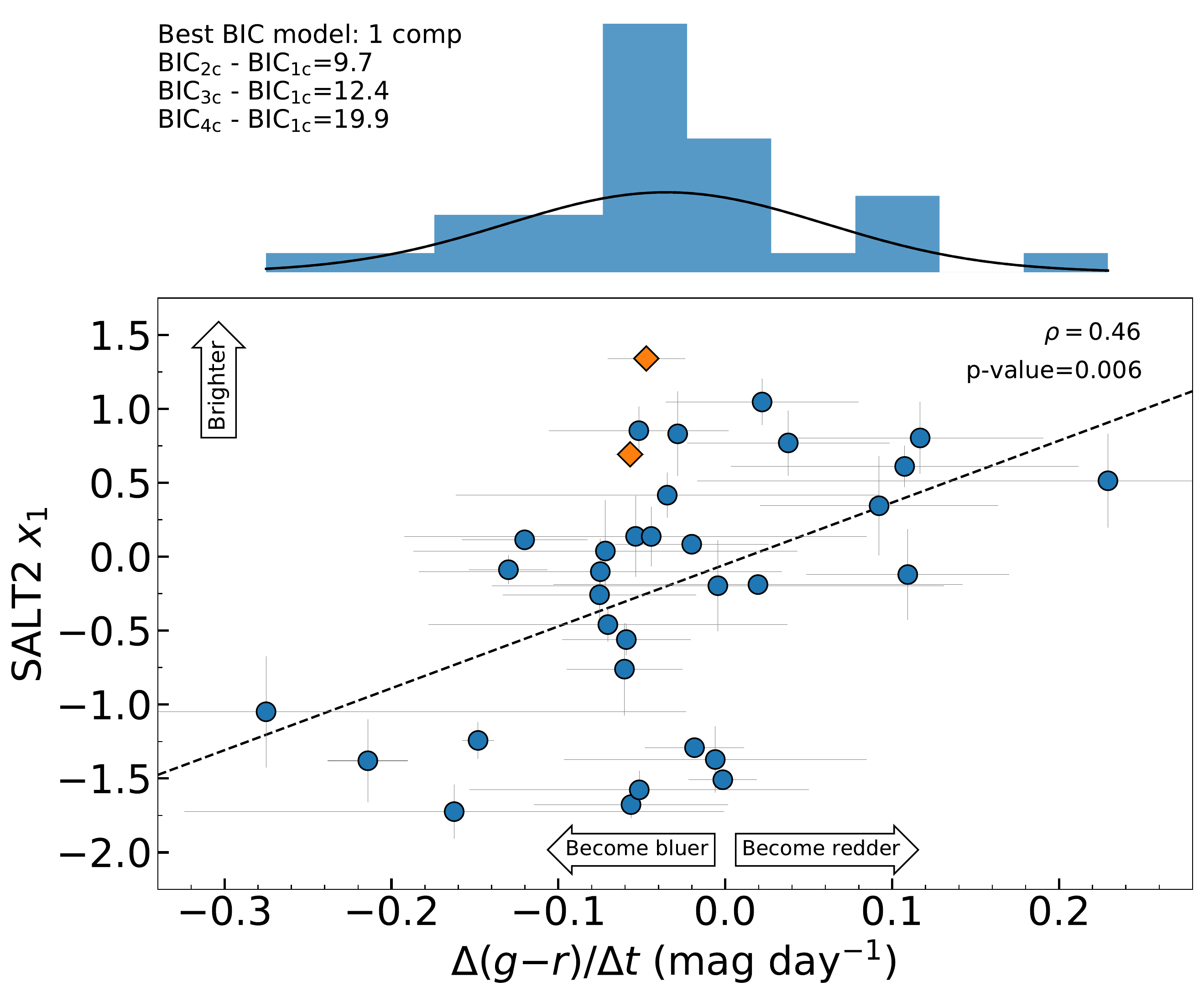}
    \caption{\textit{Top panel}: test of multiple populations in the slope distribution, using a linear fit to the slope in the first \edit1{6} days.
    %(left) and the instantaneous slope in color controlled by the difference between the best-fit power-law indexes, $\alpha_r-\alpha_g$ (right, see Equation~\ref{eq:slope}). 
    The distribution is consistent with being drawn from a single population (\referee{i.e.,} one component), with multiple components strongly disfavoured 
    %in the instantaneous slope distribution 
    ($\Delta$BIC~$>6$, see text for details). \textit{Bottom panel}: SALT2 $x_1$ parameter as a function of the linear slope \slope.
    %and the instantaneous slope $\alpha_r-\alpha_g$ (right). 
    A significant correlation is found (Pearson's coefficient $\rho=$~\edit1{0.46}, p-value=\edit1{0.006}). %No clear correlation is found between $x_1$ and $\alpha_r-\alpha_g$ although the five SNe with $\alpha_r-\alpha_g>0$ are all relatively over-luminous ($x_1\gtrsim0.7$). 
    The analysis is restricted to \edit1{34}~SNe~Ia with at least three detections in the first \edit1{6}~days (the peculiar 02cx-like SN ZTF18abclfee/SN\,2018cxk is excluded from this analysis). Symbols are the same as in Figure~\ref{fig:gr}.
   \label{fig:gr_slope} }
\end{figure*}

In this section, we take a closer look at the color evolution of $g-r$ colors at early phases, with the aim of testing the claim of two distinct populations made by \citet[][see Section~\ref{sec:intro}]{Stritzinger2018}. Specifically, we will base our discussion on \edit1{34}~SNe~Ia with reliable $g-r$ slopes in the first \edit1{6}~days (\slope) as discussed in Section~\ref{sec:evolution} (the peculiar 02cx-like SN ZTF18abclfee/SN\,2018cxk is excluded from this analysis). In particular, we will first test the presence of multiple populations in Section~\ref{sec:gaussian} and then search for possible correlations between color evolution and brightness in Section~\ref{sec:lcpar}.

%\begin{figure*}[t]
%    \centering
%    \includegraphics[width=0.85\textwidth]{plots/gr_pop.pdf}
%   \caption{Test for multiple populations in the early-time $g-r$ color evolution. The g-r slope in the first 4~days, d$(g-r)$/d$t$, is calculated for events discovered within 2~days from $t_\mathrm{fl}$ and with at least three measured colors. The distribution of d$(g-r)$/d$t$ values is shown in the top panel together with a one component model (red, best-fit model according to BIC)
   %and a two-component model (black, best-fit model according to AIC). 
%   Results of information criteria are shown in the bottom panels. The bottom-left panel refers to results using the ``fiducial sample'', with BIC values for each component shown relative to the best-fit one-component model. The horizontal dashed line at $\Delta$(IC)=6 marks the level above which a model is strongly disfavoured compared to the best-fit model. The bottom-right panel refers to results using the ``Bootstrap samples'', showing the fraction of times a model with a given number of components is preferred according to BIC. 
%   \label{fig:gr_pop}}
%\end{figure*}

\subsection{Gaussian Mixture Models}
\label{sec:gaussian}

%Given the relatively large uncertainties and moderate homogeneity of colors at early phases (Section~\ref{sec:colors}), testing the claim of distinct populations requires some statistical analysis. Specifically, w
To test the claim of distinct populations, we apply Gaussian mixture models with single or multiple components to \slope.
%~and $\Delta\alpha$. 
In order to select how many components best fit the data, we use 
%two information criteria (IC): the Akaike information criteria (AIC, \citealt{Akaike1974}) defined as
%\begin{equation}
%\mathrm{AIC} = -2\,\mathrm{ln}\mathcal{L} + 2\,k %   
%\end{equation}
the Bayesian information criteria (BIC, \citealt{Schwarz1978}) defined as
\begin{equation}
\mathrm{BIC} = -2\,\mathrm{ln}\mathcal{L} + k\,\mathrm{ln}N~~,
\end{equation}
where $\mathcal{L}$ is the maximum likelihood, $N$ the number of data points and $k$ the number of parameters. The best model is the one with the lowest BIC, with the other models strongly disfavoured if the difference to the best-fit model, $\Delta$(BIC), is larger than 6 \citep[see \referee{e.g.,} ][]{Sollerman2009}. The difference between different IC approaches lies in how much multiple-component models (and thus an added complexity) are penalised compared to a single-component model. As discussed in \citet{Liddle2004}, we choose BIC as this penalizes complexity/number of parameters more compared to \referee{e.g.,}  the Akaike information criteria (AIC, \citealt{Akaike1974}).

Results of this analysis are summarized in the top panel of Figure~\ref{fig:gr_slope}, where we show the distribution of \slope~values together with the BIC best-fit model. 
%(left) and $\Delta\alpha$~(right) 
%For the linear slope \slope, w
We find BIC~$\sim$~\edit1{$-$56.3, $-$46.6, $-$43.9, $-$36.4} for models with one, two, three and four components, respectively. Therefore, the distribution is consistent with being drawn from one single population, \referee{i.e.,} min(BIC)\,=\,BIC$_\mathrm{1C}$. In addition, a one-component is not only preferred but strongly favoured over multiple-component models ($\Delta$BIC~$>6$).
%but multiple components are not strongly disfavoured ($\Delta$BIC~$<6$). 
%For the linear slope \slope, we find BIC~$\sim-22.1, -15.8,-7.1,0.0$ for models with one, two, three and four components, respectively. 
%For the ``power-law index'' slope $\Delta\alpha$, we find instead BIC~$\sim-15.2, -7.8,+0.7,-1.7$ for one-to-four Gaussian components. That is, a one-component is not only preferred but strongly favoured over multiple-component models ($\Delta$BIC~$>6$).

To summarize, the color evolution in the first $\sim$~\edit1{6} days after first light does not show any evidence for two or multiple components and it is consistent with being drawn from a single population. This conclusion is in contrast with the claim in \citet{Stritzinger2018} although we note that $B-V$ might be a better combination compared to $g-r$ to test for the presence of multiple populations (see Section~\ref{sec:colors}). Our findings are consistent with the $B-V$ color evolution reported in \citet[][see their figure 5]{Han2020}, where there appears to be no gap between the ``red'' and ``blue'' class when adding six events to the sample of \citet{Stritzinger2018}. Surprisingly, \citet{Han2020} claims the presence of two distinct classes, although we note that similarly to 
\citet{Stritzinger2018} no analysis is provided to corroborate their conclusion.
 
%, while some marginal evidence for multiple populations is found when AIC is used. Given that BIC is more penalizing to added complexity in the model, we conservatively favour one single component to explain the $g-r$ evolution of our sample in the first 4~days after $t_\mathrm{fl}$. 
%We note, however, that the relatively small scatter and large uncertainty at early phases (see Figure~\ref{fig:gr_stritz}) means even two distinct populations might overlap in the \slope~parameter space and thus be hard to distinguish.

\subsection{Color evolution vs brightness}
\label{sec:lcpar}

The bottom panel of Figure~\ref{fig:gr_slope} show values of \slope~against the SALT2 $x_1$ parameter, %~(left) and $\Delta\alpha$~(right) 
where the latter is used as a proxy for the SN brightness (with brighter events corresponding to larger $x_1$). We find a moderate correlation between the linear slope \slope~and SALT2 $x_1$. Specifically, the Pearson's correlation coefficient of $\rho=$~\edit1{0.46} suggests that this correlation is significant (p-value of \edit1{0.006}, \referee{i.e.,} statistically significant at the significance level of 0.01). Relatively brighter events (large $x_1$) are preferentially associated to \mbox{$g-r$} colors that are flat or evolving to redder colors, \slope~$\gtrsim$~0. In contrast, relatively fainter events (small $x_1$) are characterized by colors becoming bluer with time, \slope~$<$~0. \edit1{We have tested, and confirmed, that this correlation is present and statistically significant for different choices of the selection criterion, i.e., the correlation persists for cuts at 3, 3.5, 4 and 4.5 days since first light.}
%We note that the correlation is stronger when restricting to events with an uncertainty on \slope~smaller than 0.05 mag day$^{-1}$, with a Pearson coefficient of $\rho=0.79$ (p-value~$<10^{-5}$).

%No correlation is instead found between $\Delta\alpha$ and SALT2 $x_1$ ($\rho=0.12$). However, the relatively large uncertainties on $\alpha_g$ and $\alpha_r$ translate to error bars of $\sim$0.2 in $\Delta\alpha$, i.e. about a third of the  $\Delta\alpha$ dispersion inferred for the sample, thus possibly hiding any correlation between $\Delta\alpha$ and $x_1$. Nevertheless, we note that the five events with $\Delta\alpha>0$ are over-luminous ($x_1\gtrsim0.7$).

To summarize, we find a moderate correlation between brightness and color slope in the first $\sim$~\edit1{6} days, with brighter events preferentially associated to flatter evolutions while fainter SNe characterized by a transition from redder to bluer colors. These two behaviours are in qualitative agreement with those identified by \citet{Stritzinger2018} for their ``blue'' and ``red'' classes, respectively. However, our findings suggests that these are only the extremes of a continuous behaviour, thus arguing against a bimodality \citep[][see also Section~\ref{sec:gaussian}]{Stritzinger2018}.

\section{Discussion and conclusions} \label{sec:concl}

We presented $g-r$ colors for a sample of \edit1{65}~SNe~Ia discovered within \edit1{5}~days from first light by ZTF in 2018. The size of our sample is \edit1{about three times larger than} the one available in the literature and extends to higher redshifts (up to $z=$~\edit1{0.143}). We find that \mbox{$g-r$} colors are relatively homogeneous at all the phases investigated, from first light to $\sim$~two weeks after. In particular, the observed scatter of $\sim$~1.5 mag at very early phases ($\lesssim$~\edit1{6}~days) is roughly half intrinsic and half due to high photometric uncertainties. Specifically, we find that the intrinsic dispersion in $g-r$ colors in the first few days after explosion is smaller than that found in $B-V$ colors \citep{Stritzinger2018} as a consequence of the different wavelength regions probed by different filter combination \citep{Nordin2018}.

We do, however, note different behaviours in the color \textit{evolution} from first light to $\sim$~\edit1{6} days later. In particular, some events have a rather steep change from redder to bluer colors while others are characterized by a flatter evolution. We further identify a significant correlation ($\rho=$~\edit1{0.46}, p-value of \edit1{0.006}) between the SALT2 $x_1$ parameter and the linear color slope in the first \edit1{6}~days, indicating that brighter events (large $x_1$) have flatter color evolutions at early times. However, contrary to previous claims in the literature \citep{Stritzinger2018}, the slope distribution does not show any evidence for bimodality and it is consistent with being drawn from a single population. \referee{We note that our findings are based on a sub-sample of \edit1{34} normal SNe~Ia with at least three detections in the first \edit1{6} days since first light, a sample that is about \edit1{twice (and not three times, see above) as large as} the one in the literature after applying the same criteria \citep{Stritzinger2018,Han2020}.}

The range in early-time slopes is reminiscent of mixing models, where an increasing amount of $^{56}$Ni mixing in the outer ejecta regions leads to a transition from colors rapidly changing from red to blue to colors with a flatter evolution. In this context, the correlation found between early-time color slopes and brightness suggests that stronger mixing (hence flatter color evolution) might occur in explosions producing more $^{56}$Ni (hence brigther). At the same time, the range in early-time slopes is in good agreement with predictions from helium-ignited ``double-detonation'' models with very thin helium layers ($M_\mathrm{He}=0.01\,M_\odot$) and varying carbon-oxygen masses between $0.9$ and $1.2\,M_\odot$ \citep{Polin2019a}. In addition, \edit1{six}~SNe in our sample show evidence for a distinctive early-time ``red bump'' predicted by ``double-detonation'' models with larger helium masses \citep[$0.02\,M_\odot\lesssim M_\mathrm{He}\lesssim0.07\,M_\odot$,][]{Noebauer2017,Polin2019a}. Our findings support recent claims in the literature arguing that a subset of SNe~Ia originates from ``double-detonation'' explosions \citep{Cikota2019,Polin2019a,Polin2019b}. In contrast, we find no clear evidence for a rapid transition from blue to red colors predicted by the ejecta-companion model discussed by \citet{Kasen2010}, posing serious challenges to this scenario for explaining the bulk of SNe~Ia.

%Finally, a reasonable agreement is found for pulsation delayed-detonation models \citep{Dessart2014}, especially with objects in our sample that have blue colors and relatively flat evolution in $g-r$.

%We find marginal evidence for two populations when inspecting the color evolution in the first 4~days more closely. Specifically, events that show a steep decline and are consistent with relatively small amount of mixing are preferentially under-luminous. Events with a flatter color evolution and more consistent with strong mixing are instead preferentially over-luminous. This tentative association is compatible with earlier suggestions from \citet[][see e.g. their figure 3]{Stritzinger2018} and \citet{Jiang2018}. 

%These associations, together with the lack of a clear correlation between brightness and color slopes, are suggestive of two distinct channels giving rise to the observed SN~Ia diversity. If confirmed, our findings would bring additional support to studies \citep[e.g.][]{Fisher2015,Blondin2017,Dhawan2018,Shen2018} suggesting that the bright-end of the SN Ia population could be explained by M$_\mathrm{ch}$ SD scenarios, in which strong ejecta mixing and thus bluer and flatter color evolution are expected, while the faint-end explained by a different channel (perhaps sub-M$_\mathrm{ch}$) producing less mixing in the ejecta and thus redder and more-rapidly evolving colors at early phases. 

Based on the number of young SNe~Ia discovered from May to December 2018 and presented here, the 3-year ZTF survey is expected to have a final sample of at least $\sim$~\edit1{200} SNe~Ia discovered within \edit1{5} days from first light. Such a large sample will allow us to place stronger constraints on explosion models and test the possible correlation between color evolution and brightness identified in this work.

%{\comment Correlation between colour and other photometric properties (e.g. Bmax or dm15)? Correlation between colour and host galaxy types (check those with NED host redshift)? }
%{\comment Go back to plot with subtypes. Are 91T-like preferentially bluer? Maybe consistent with more mixing seen in overluminous SNe (16abc, 12fr etc, citations). Cite also Fisher's 2015 paper on connection between SD (off-set deflagration) and 91T/overluminous SNe. Also, claim by Blondin and Ken shen that SD/Mch might explain brighter end of brightness distribution. What about  02cx-like? Is this consistent with what seen in the literature? Check Saurabh's 2017 review paper on 02cx-like.}

%\section{Comparison to templates} \label{subsec:templ}

%{\comment Overplot light curve templates (11fe, Hsiao, Nugent). How good they are? Show residuals. Perhaps they are good later on (from roughly 5 days after explosion) but not in the early stages.}

%{\comment Propose our template. Gaussian processes by Semeli.}

%{\comment Wrap this up.}

%\begin{figure*}[t]
%    \centering
%    \includegraphics[width=\textwidth]{plots/gr_pop.pdf}
%   \caption{\comment{Caption}. \label{fig:gr_pop}}
%\end{figure*}

\acknowledgments

The authors are thankful to Tony Piro for sharing his models, and to Chris Ashall, Joel Johansson, Mark Magee, Keiichi Maeda and Stuart Sim for useful discussions. 

MB acknowledges support from the G.R.E.A.T research environment funded by the Swedish National Science Foundation. A.A.M.~is funded by the Large Synoptic Survey Telescope Corporation, the
Brinson Foundation, and the Moore Foundation in support of the LSSTC Data
Science Fellowship Program; he also receives support as a CIERA Fellow by the
CIERA Postdoctoral Fellowship Program (Center for Interdisciplinary
Exploration and Research in Astrophysics, Northwestern University). This research was supported in part
through the computational resources and staff contributions provided for the
Quest high performance computing facility at Northwestern University which is
jointly supported by the Office of the Provost, the Office for Research, and
Northwestern University Information Technology. This work was supported in
part by the GROWTH project funded by the National Science Foundation under
Grant No.~1545949. SRK thanks the Heising-Simons Foundation for supporting his ZTF research.

This work is based on observations obtained with the Samuel Oschin Telescope 48-inch and the 60-inch Telescope at the Palomar Observatory as part of the Zwicky Transient Facility project. ZTF is supported by the National Science Foundation under Grant No. AST-1440341 and a collaboration including Caltech, IPAC, the Weizmann Institute for Science, the Oskar Klein Center at Stockholm University, the University of Maryland, the University of Washington, Deutsches Elektronen-Synchrotron and Humboldt University, Los Alamos National Laboratories, the TANGO Consortium of Taiwan, the University of Wisconsin at Milwaukee, and Lawrence Berkeley National Laboratories. This work was supported by the GROWTH project \citep{Kasliwal2019} funded by the National Science Foundation under Grant No 1545949. Operations are conducted by COO, IPAC, and UW. This work made use of the Heidelberg Supernova Model Archive (HESMA), \url{https://hesma.h-its.org}. 

%\appendix

%\renewcommand\thefigure{\thesection\arabic{figure}} 

%\section{Color evolution of individual supernovae} \label{sec:individual}

%\setcounter{figure}{0}

%In Figure~\ref{fig:gr_ind}, we highlight the $g-r$ color evolution of all the 69~SNe in the sample ordered by SALT2 $x_1$.

\bibliography{bulla2020a}
\bibliographystyle{aasjournal}

\end{document}